\documentclass[conference,letterpaper]{IEEEtran}
\IEEEoverridecommandlockouts 

\usepackage[utf8]{inputenc} 
\usepackage[T1]{fontenc} 
\usepackage[cmex10]{amsmath} 

\usepackage{amssymb} 
\usepackage{mathtools} 
\usepackage{amsfonts} 
\usepackage{accents} 

\usepackage{mleftright}\mleftright 

\usepackage{multirow}

\usepackage[dvips]{graphicx} 
\graphicspath{{./}}
\DeclareGraphicsExtensions{.eps}

\usepackage[dvipsnames]{xcolor} 

\usepackage{bbm} 
\usepackage{xfrac} 
\usepackage{cancel} 
\usepackage{wrapfig} 
\usepackage{tensor} 

\usepackage{theorem}
\usepackage{cite}

\usepackage{url} 

\usepackage{psfrag} 

\usepackage[inline]{enumitem} 

\usepackage{array} 

\usepackage{aliascnt} 

\usepackage[hidelinks,linktocpage]{hyperref}

\usepackage{cleveref} 

\usepackage[ruled,algoruled,vlined]{algorithm2e}

\SetCommentSty{mycommfont}

\usepackage{xparse} 

\usepackage{ifthen} 

\usepackage{comment}

\crefname{section}{Section}{Sections}
\crefname{subsection}{Section}{Sections}
\crefname{equation}{Eq.}{Equations}
\crefname{enumi}{part}{parts}
\crefname{table}{Table}{Tables}
\crefname{figure}{Figure}{Figures}
\crefname{algocf}{Algorithm}{Algorithms}

\newtheorem{theorem}{Theorem}
\crefname{theorem}{Theorem}{Theorems}

\newaliascnt{lemma}{theorem}
\newtheorem{lemma}[lemma]{Lemma}
\aliascntresetthe{lemma}
\crefname{lemma}{Lemma}{Lemmas}

\newaliascnt{definition}{theorem}
\newtheorem{definition}[definition]{Definition}
\aliascntresetthe{definition}
\crefname{definition}{Definition}{Definitions}

\newaliascnt{corollary}{theorem}
\newtheorem{corollary}[corollary]{Corollary}
\aliascntresetthe{corollary}
\crefname{corollary}{Corollary}{Corollarys}

\newaliascnt{claim}{theorem}

\aliascntresetthe{claim}
\crefname{claim}{Claim}{Claims}

\newaliascnt{conjecture}{theorem}
\newtheorem{conjecture}[conjecture]{Conjecture}
\aliascntresetthe{conjecture}
\crefname{conjecture}{Conjecture}{Conjectures}

\newaliascnt{question}{theorem}

\aliascntresetthe{question}
\crefname{question}{Question}{Questions}

\newaliascnt{oquestion}{theorem}

\aliascntresetthe{oquestion}
\crefname{oquestion}{Open Question}{Open Questions}

\newtheorem{examplex}{Example}

\theoremstyle{plain}
\theorembodyfont{\normalfont}

\newtheorem{cnstr}{Construction}

\crefname{cnstr}{Construction}{Constructions}

\crefname{step}{Step}{Steps}

\crefname{regime}{Regime}{Regimes}

\newtheorem{myalgo}{Algorithm}

\crefname{myalgo}{Algorithm}{Algorithms}

\newcommand\numberthis{\stepcounter{equation}\tag{\theequation}}

\newcounter{enumrom}
\renewcommand{\theenumrom}{(\roman{enumrom})}

\makeatletter
\renewcommand{\@endtheorem}{\endtrivlist}
\makeatother

\makeatletter
\renewcommand{\thefigure}{{\@arabic\c@figure}}
\renewcommand{\fnum@figure}{{\bf Figure\,\thefigure}}
\makeatother

\renewcommand{\leq}{\leqslant}
\renewcommand{\geq}{\geqslant}

\newcommand{\cA}{\mathcal{A}}

\newcommand{\cQ}{\mathcal{Q}}

\renewcommand{\Bbb}{\mathbb}

\newcommand{\bbC}{{\Bbb C}}
\newcommand{\bbN}{{\Bbb N}}
\newcommand{\bbR}{{\Bbb R}}
\newcommand{\bbZ}{{\Bbb Z}}
\newcommand{\bbE}{{\Bbb E}}

\newcommand{\Z}{{\Bbb Z}}

\newcommand{\1}{\mathbbm{1}}

\newcommand{\awcomment}[1]{\textcolor{blue}{#1}}
\newcommand{\hacomment}[1]{\textcolor{blue!50!black!50!green}{#1}}

\DeclarePairedDelimiter\abs{\lvert}{\rvert}
\DeclarePairedDelimiter\ceilenv{\lceil}{\rceil}
\DeclarePairedDelimiter\floorenv{\lfloor}{\rfloor}
\DeclarePairedDelimiter\parenv{\lparen}{\rparen}
\DeclarePairedDelimiter\sparenv{\lbrack}{\rbrack}
\DeclarePairedDelimiter\bracenv{\lbrace}{\rbrace}
\DeclarePairedDelimiterX\mathset[2]{\lbrace}{\rbrace}{#1 : #2}
\DeclarePairedDelimiterX\inner[2]{\langle}{\rangle}{#1 \mathrel{},\mathrel{} #2}
\DeclarePairedDelimiterX\condparenv[2]{(}{)}{#1 \mathrel{}\delimsize\vert\mathrel{} #2}

\DeclareDocumentCommand\norm{ o m }{
    \IfNoValueTF{#1}
        {\left\Vert#2\right\Vert}
        {\left\Vert#2\right\Vert_{#1}}
}

\DeclareDocumentCommand\der{ o m o }{
    \IfNoValueTF{#1}
        {
            \IfNoValueTF{#3}
                {\frac{d}{d{#2}}}
                {\frac{d{#3}}{d{#2}}}
        }
        {\parenv*{\frac{d}{d{#2}}}^{#1}\IfNoValueTF{#3}{}{#3}}
}
\DeclareDocumentCommand\partder{ o m m }{
    \IfNoValueTF{#1}
        {\frac{\partial{#3}}{\partial{#2}}}
        {\frac{\partial^{#1}{#3}}{{\partial{#2}}^{#1}}}
}
\DeclareDocumentCommand\df{ o m o }{
    d\IfNoValueTF{#1}{}{^{#1}}{#2}\IfNoValueTF{#3}{}{_{#3}}
}

\newcommand{\deq}{\mathrel{\triangleq}}

\newcommand{\qa}{\hat{q}_{\mathtt{a}}}
\newcommand{\qg}{\hat{q}_{\mathtt{g}}}
\newcommand{\qmg}{\hat{q}_{\mathtt{mg}}}
\newcommand{\qmh}{\hat{q}_{\mathtt{mh}}}

\def\ve#1{{%
        \mathchoice{\mbox{\boldmath$\displaystyle #1$}}%
		{\mbox{\boldmath$\textstyle #1$}}%
		{\mbox{\boldmath$\scriptstyle #1$}}%
		{\mbox{\boldmath$\scriptscriptstyle #1$}}%
}}

\newcommand{\lB}{\lambda_B}
\newcommand{\fB}{f_B}

\DeclareFontFamily{U}{mathx}{\hyphenchar\font45}
\DeclareFontShape{U}{mathx}{m}{n}{
      <5> <6> <7> <8> <9> <10>
      <10.95> <12> <14.4> <17.28> <20.74> <24.88>
      mathx10
      }{}
\DeclareSymbolFont{mathx}{U}{mathx}{m}{n}
\DeclareFontSubstitution{U}{mathx}{m}{n}
\DeclareMathAccent{\widecheck}{0}{mathx}{"71}
\DeclareMathAccent{\wideparen}{0}{mathx}{"75}

\DeclareMathOperator{\supp}{supp}

\DeclareMathOperator{\wt}{wt}

\DeclareDocumentCommand\enc{ o }{
    \IfNoValueTF{#1}
        {\operatorname{Enc}}
        {\operatorname{Enc}_{\ref*{#1}}}
}

\DeclareDocumentCommand\dec{ o }{
    \IfNoValueTF{#1}
        {\operatorname{Dec}}
        {\operatorname{Dec}_{\ref*{#1}}}
}

\begin{document}

\title{Bounds on Mixed Codes with Finite Alphabets}

\author{%
    \IEEEauthorblockN{\textbf{Yonatan~Yehezkeally}\IEEEauthorrefmark{1}, 
                     \textbf{Haider~Al~Kim}\IEEEauthorrefmark{1}\IEEEauthorrefmark{2}, 
                     \textbf{Sven~Puchinger}\IEEEauthorrefmark{3}, 
                     and \textbf{Antonia~Wachter-Zeh}\IEEEauthorrefmark{1}}
   \IEEEauthorblockA{\IEEEauthorrefmark{1}%
                     School of Computation, Information and Technology, 
                     Technical University of Munich, 
                     80333 Munich, Germany}
   \IEEEauthorblockA{\IEEEauthorrefmark{2}%
                     Department of Electrical and Communication Engineering, 
                     University of Kufa, Iraq}
   \IEEEauthorblockA{\IEEEauthorrefmark{3}%
                     Hensoldt Sensors GmbH, 
                     89077 Ulm, Germany}
    \thanks{%
        This project has received funding from the European Research Council (ERC) under the European Union's Horizon 2020 research and innovation programme (grant agreement No. 801434).
        Y. Yehezkeally was supported by a Carl Friedrich von Siemens postdoctoral research fellowship of the Alexander von Humboldt Foundation.}%
}
\maketitle
\makeatother
\begin{abstract}
Mixed codes, which are error-correcting codes in the 
Cartesian product of different-sized spaces, model degrading storage 
systems well. While such codes have previously been studied for their 
algebraic properties (e.g., existence of perfect codes) or in the case 
of unbounded alphabet sizes, we focus on the case of finite alphabets, 
and generalize the Gilbert-Varshamov, sphere-packing, Elias-Bassalygo, 
and first linear programming bounds to that setting. 
In the latter case, our proof is also the first for the non-symmetric 
mono-alphabetic $q$-ary case using Navon and Samorodnitsky's 
Fourier-analytic approach.
\end{abstract}

\section{Introduction} \label{sec:intro}

In traditional coding theory, one generally studies codes where every 
coordinate is over the same alphabet (e.g., binary--or, more 
generally, $q$-ary--codes). A rich body of knowledge has been 
developed regarding constructions, and bounds on the parameters, of 
such error-correcting codes.

However, in some circumstances this assumption might not hold. 
Sidorenko et al. \cite{SidSchGabBosAfa05} suggested that this is the 
case in orthogonal-frequency-division-multiplexing (OFDM) 
transmission; it can also be viewed as a relaxation of the 
partially-stuck-cell setting \cite{AlKPucWac19}, where both sender and 
receiver are aware (perhaps thorough a periodic sampling routine) of 
which coordinates have smaller alphabets. The authors further believe 
that this generalization of classical error correction is of 
independent theoretical interest (see, e.g., their study 
in~\cite[Ch.~7]{Etz22}).

Codes designed for the setting of varying alphabet sizes are named 
\emph{mixed}- (or \emph{polyalphabetic}-)\linebreak codes, and have 
been studied in the past. In comparison to \cite{Sch70, HerSch71, 
HerSch72, Lin75, Hed77, EtzGre93}, the codes we consider are not 
necessarily perfect, and therefore can correct more than a single 
error. On the other hand, \cite{SidSchGabBosAfa05} generalized the 
Singleton bound to mixed codes, and presented constructions of 
MDS codes, based on this bound and known MDS ``mother codes'', by 
letting alphabet sizes grow with respect to the code block size. 
Similarly, \cite[Cha.~7.3]{Etz22} constructed diameter-perfect mixed 
codes that meet a code-anticode bound proven therein, which also rely 
on unbounded alphabet sizes. In contrast to both, we study the setting 
where alphabet sizes are bounded.

The rest of this manuscript is organized as follows. In 
\cref{sec:contr}, we summarize the main contributions of this work. In 
\cref{sec:prelim} we present definitions and notations. Then, in 
\cref{sec:sphere} we study the size of Hamming spheres in this space; 
in \cref{sec:list} we observe the list-decoding capabilities of mixed 
codes by generalizing the first Johnson bound, and in 
\cref{sec:bounds} we develop lower- and upper bounds on the sizes of 
mixed codes. Finally, in \cref{sec:conc} we demonstrate that our 
bounds improve upon the known bound 
of~\cite[Th.~2]{SidSchGabBosAfa05},\cite[Cor.~2.15]{Etz22} in some 
settings.

\section{Main contribution} \label{sec:contr}

Our contributions in this work are as follows: 
\begin{enumerate}[label=(\roman*)]
\item
While \cite{SidSchGabBosAfa05} presented a Gilbert-Varshamov lower 
bound and a sphere-packing upper bound based on a straight-forward 
expression for sphere size, containing an exponential number of terms, 
we develop a recursive formula for the size of spheres which enables 
one to efficiently compute exact sizes in any given case, resulting in 
said bounds on code sizes.

\item
We develop closed-form upper and lower bounds on the size of spheres, 
yielding asymptotic expressions for the size of balls which readily 
lend themselves to closed-form statements of the asymptotic 
Gilbert-Varshamov and sphere-packing bounds (more precisely, lower and 
upper bounds on these, respectively).

\item
In comparison to a known bound (\hspace{1sp}%
\cite[Th.~2]{SidSchGabBosAfa05} and \cite[Cor.~2.15]{Etz22}, which 
curiously develop the same bound in this context), we develop the 
equivalence of the Elias-Bassalygo bound (in~\cite{Bas65b}, and 
reported in~\cite{Joh63}) and the first linear-programming (LP) 
bound~\cite{McERodRumWel77, Lev98} for mixed codes, which we show are 
tighter, for codes with some minimum distances, when alphabet sizes 
are bounded. In particular, our treatment of the LP~bound relies on 
Navon and Samorodnitsky's Fourier analysis approach~\cite{NavSam09}; 
to our knowledge its restriction to fixed alphabet size is the first 
time that a proof for the bound in the general $q$-ary case (e.g., not 
only symmetric codes) is suggested using these methods.

\end{enumerate}

\section{Preliminaries} \label{sec:prelim}

The pertinent space is defined as follows. For $n\in\bbN$, take some $1<q_1\leq q_2\leq\ldots\leq q_n$. For convenience we denote $[n] = \bracenv*{1,2,\ldots,n}$. We let $\cQ\deq \prod_{i=1}^n (\bbZ/q_i\bbZ)$. This Cartesian product should be interpreted as a (finite) product of (finite) cyclic groups, with the resulting structure of a finite Abelian group.

We endow $\cQ$ with the Hamming metric, defined $d(x,y)\deq \wt(x-y)$, where $\wt(x)\deq\abs*{\supp(x)}$. 
We denote the \emph{sphere of radius $r$ around $x\in\cQ$} in this metric by $S_r(x)\deq \mathset*{y\in\cQ}{d(x,y)=r}$, and the \emph{ball of radius $r$ around $x\in\cQ$} by $B_r(x)\deq \mathset{y\in\cQ}{d(x,y)\leq r}$. 
We say a \emph{code} $C\subseteq\cQ$ has \emph{minimum distance}~$d$ if for all $x,y\in C$, $x\neq y$ implies $d(x,y)\geq d$. 
Let $A(n,d),A_r(n,d),A_{\leq r}(n,d)$ denote the maximum size of code 
in $\cQ,S_r(0),B_r(0)$ respectively, with minimal Hamming 
distance~$d$.

The \emph{rate} of a code $C\subseteq\cQ$ is defined by $R(C)\deq \frac{\log\abs*{C}}{\log\abs*{\cQ}}$. For $0\leq\delta\leq 1$ we shall be interested in the \emph{maximum achievable rate} 
\begin{align}
    R(\delta) 
    &\deq \frac{\log A(n,\floorenv*{\delta n})}{\log\abs*{\cQ}}.
\end{align}
We will also be interested in an asymptotic analysis of $R(\delta)$ 
as $n$ grows to infinity; in these cases, we shall assume 
$\bracenv*{q_i}$ is sampled from a fixed set of alphabet sizes, where 
the incidence of each value is proportional to $n$.

Finally, we make the following notation for the arithmetic and 
geometric means of the alphabet sizes~$\bracenv*{q_i}_{i=1}^n$:
\begin{align}
    \qa \deq \frac{1}{n}\sum_{i=1}^n q_i; \qquad
    \qg \deq \parenv*{\prod_{i=1}^n q_i}^{1/n},
\end{align}
as well as geometric and harmonic means of $\bracenv*{q_i-1}_{i=1}^n$:
\begin{align}
    \qmg \deq \parenv*{\prod_{i=1}^n (q_i-1)}^{\;\;\mathclap{1/n}}+1; 
    \qquad \qmh \deq \frac{n}{\sum_{i=1}^n \frac{1}{q_i-1}} + 1.
\end{align}
(Note that in our notation $\qmg-1, \qmh-1$ are the geometric and 
harmonic means of $\bracenv*{q_i-1}_{i=1}^n$, respectively; further, 
$\qa-1$ is its arithmetic mean.)

Observe that $\qa\geq \qg\geq \qmg\geq \qmh$. Further, if $n$ grows 
and $\bracenv*{q_i}_{i=1}^n$ is sampled as described above, $\qa, \qg, 
\qmg, \qmh$ are fixed with respect to $n$.

\section{Size of Spheres} \label{sec:sphere}

In this section, we study the size of spheres $S_r(x)$ in Hamming 
distance on $\cQ$. Since the Hamming distance is shift-invariant, we 
denote the size of the Hamming sphere $s_r\deq \abs*{S_r(0)}$, for 
$0\leq r\leq n$. 
It was noted in \cite[Eq.~5]{SidSchGabBosAfa05} that 
$
s_r = \sum_{1\leq i_1<i_2<\ldots<i_r\leq n} \prod_{j=1}^r (q_{i_j}-1)
$; 
however, this expression contains an exponential number of summands, and is challenging to work with. In this section, we instead develop a recursive expression for $s_r$, which can be evaluated in polynomial time, as well as develop bounds on it.

For convenience, we also denote for all $I\subseteq [n]$, $S_r(I)\deq 
\mathset*{x\in S_r(0)}{\supp(x)\subseteq I}$ and $s_r(I)\deq 
\abs*{S_r(I)}$ (so that $s_r = s_r([n])$). Then, observe the 
following: 
\begin{lemma}\label{lem:sum-sphere-not-supported}
\begin{enumerate}
\item\label{par:sum-sphere-not-supported-1}
For all $0\leq r\leq n$, $\sum_{i\in [n]} 
s_r([n]\setminus\bracenv*{i}) = (n-r) s_r$.

\item\label{par:sum-sphere-not-supported-2}
For all $0\leq r < n$, $\sum_{i\in [n]} (q_i-1) 
s_r([n]\setminus\bracenv*{i}) = (r+1) s_{r+1}$.

\item\label{par:sum-sphere-not-supported-3}
For all $0\leq r < n-1$, $\sum_{i\in [n]} (q_i-1)^2 
s_r([n]\setminus\bracenv*{i}) = n(\qa-1) s_{r+1} - (r+2) s_{r+2}$.

\end{enumerate}
\end{lemma}
\begin{IEEEproof}
\begin{enumerate}
\item
Observe for $x\in S_r(0)$ that $x\in S_r([n]\setminus\bracenv*{i})$ if and only if $i\in\supp(x)$.

\item
We note 
\begin{align*}
	(q_i-1) s_r([n]\setminus\bracenv*{i}) 
	= \abs*{\mathset*{x\in S_{r+1}(0)}{i\in\supp(x)}}.
\end{align*}

\item
Observe 
\begin{IEEEeqnarray*}{+rCl+x*}
	\IEEEeqnarraymulticol{3}{l}{\sum_{i\in [n]} (q_i-1)^2 
	s_r([n]\setminus\bracenv*{i})} \\
	\;\; &=& n(\qa-1) s_{r+1} \\
	&&\>- \sum_{i\in [n]} (q_i-1) \parenv*{s_{r+1} - (q_i-1) s_r([n]\setminus\bracenv*{i})} \\
	&=& n(\qa-1) s_{r+1} - \sum_{i\in [n]} (q_i-1) s_{r+1}([n]\setminus\bracenv*{i}) \\
	&=& n(\qa-1) s_{r+1} - (r+2) s_{r+2}.
	\\[-\normalbaselineskip] &&&\IEEEQEDhere
\end{IEEEeqnarray*}

\end{enumerate}
\end{IEEEproof}

\begin{theorem}\label{thm:sphere-size-sum}
It holds that $s_0 = 1$, and for $0<r\leq n$, 
\begin{align*}
	s_r = \frac{1}{r} \sum_{k=0}^{r-1} (-1)^k s_{r-1-k} 
	\sum_{i\in [n]} (q_i-1)^{k+1}.
\end{align*}
\end{theorem}
\begin{IEEEproof}
That $s_0 = 1$ is immediate. Then, for $0<r\leq n$ and $1\leq i\leq n$, we note 
\begin{IEEEeqnarray*}{+rCl+x*}
	s_r([n]\setminus\bracenv*{i}) 
	&=& s_r([n]) - (q_i-1) s_{r-1}([n]\setminus\bracenv*{i}) \\
	&=& s_r([n]) - (q_i-1) 
	\big(s_{r-1}([n]) \\
	&&\>- (q_i-1) s_{r-2}([n]\setminus\bracenv*{i})\big) \\
	&=& \ldots = \sum_{k=0}^r (-1)^k (q_i-1)^k s_{r-k}.
\end{IEEEeqnarray*}
Hence from \cref{par:sum-sphere-not-supported-2} of 
\cref{lem:sum-sphere-not-supported}, 
\begin{IEEEeqnarray*}{+rCl+x*}
	s_r 
	&=& \frac{1}{r} \sum_{i\in [n]} (q_i-1) s_{r-1}([n]\setminus\bracenv*{i}) \\
	&=& \frac{1}{r} \sum_{i\in [n]} 
	\sum_{k=0}^{r-1} (-1)^k (q_i-1)^{k+1} s_{r-1-k} \\
	&=& \frac{1}{r} \sum_{k=0}^{r-1} (-1)^k s_{r-1-k} 
	\sum_{i\in [n]} (q_i-1)^{k+1}.
	\\[-\normalbaselineskip] &&&\IEEEQEDhere
\end{IEEEeqnarray*}
\end{IEEEproof}

Observe that \cref{thm:sphere-size-sum} suggests a polynomial-run-time 
algorithm for computing $s_r$ and $\abs*{B_r(x)}=\sum_{k=0}^r s_k$.

\begin{theorem}\label{thm:sphere-ratio-ave}
For $0\leq r < n$, the ratio $\frac{(r+1) s_{r+1}}{(n-r) s_r}$ is decreasing in $r$; in particular, 
\begin{align*}
	\frac{(\qmh-1) (n-r)}{r+1} \leq 
	\frac{s_{r+1}}{s_r} \leq 
	\frac{(\qa-1) (n-r)}{r+1}.
\end{align*}
\end{theorem}
\begin{IEEEproof}
Firstly, observe that $\frac{s_1}{s_0} = n (\qa-1)$ and 
$\frac{s_n}{s_{n-1}} = \frac{\qmh-1}{n}$, achieving the upper and 
lower bounds, respectively. Hence the latter part of the claim follows 
from the former.

Next, substitute in the sequel $a_i\deq q_i-1$ and $b_i\deq 
s_r([n]\setminus\bracenv*{i})$, and observe that $a_i$ ($b_i$) is 
monotone non-decreasing (non-increasing, respectively). Assume to 
the contrary that $\frac{(r+2) s_{r+2}}{(n-r-1) s_{r+1}} > 
\frac{(r+1) s_{r+1}}{(n-r) s_r}$ for some $0\leq r\leq n-2$; it 
follows that 
\begin{align*}
	(n-r) (r+1) (r+2) s_{r+2} s_r > (n-r-1) (r+1)^2 s_{r+1}^2,
\end{align*}
and from \cref{par:sum-sphere-not-supported-1,par:sum-sphere-not-supported-2,par:sum-sphere-not-supported-3} 
of \cref{lem:sum-sphere-not-supported} this is equivalent to 
\begin{IEEEeqnarray*}{+rCl+x*}
	\IEEEeqnarraymulticol{3}{l}{\parenv*{\sum_{i\in[n]} b_i} 
	\parenv*{\parenv*{\sum_{i\in[n]} a_i} 
	\parenv*{\sum_{i\in [n]} a_i b_i} 
	- (r+1) \sum_{i\in [n]} a_i^2 b_i}} \\
	\;\; &>& (n-r-1) \parenv*{\sum_{i\in[n]} a_i b_i}^2.
\end{IEEEeqnarray*}
Rearranging, we have 
\begin{IEEEeqnarray*}{+rCl+x*}
	\IEEEeqnarraymulticol{3}{l}{\parenv*{\sum_{i,j\in [n]} a_i b_i b_j 
	\parenv*{\parenv*{\sum{}_{k\in[n]} a_k} - n a_j}}} \\
	\;\; &>& (r+1) \parenv*{\sum_{i,j\in [n]} a_i b_i b_j (a_i - a_j)}.
\end{IEEEeqnarray*}
Observe that the right-hand side is non-negative, since 
\begin{IEEEeqnarray*}{+rCl+x*}
	\IEEEeqnarraymulticol{3}{l}{\sum_{i,j\in [n]} a_i b_i b_j (a_i - a_j)} \\
	\;\; &=& \frac{1}{2} \sum_{i,j\in [n]} \parenv*{a_i b_i b_j 
	(a_i - a_j) + a_j b_j b_i (a_j - a_i)} \\
	&=& \frac{1}{2} \sum_{i,j\in [n]} b_i b_j (a_i - a_j)^2 \geq 0.
\end{IEEEeqnarray*}
Hence, in particular, 
\begin{IEEEeqnarray*}{+rCl+x*}
	\IEEEeqnarraymulticol{3}{l}{\sum_{i,j\in [n]} a_i b_i b_j 
	\parenv*{\parenv*{\sum{}_{k\in[n]} a_k} - n a_j}} \\
	\;\; &>& \sum_{i,j\in [n]} a_i b_i b_j (a_i - a_j),
\end{IEEEeqnarray*}
which we rearrange to
\begin{align*}
	0 &< \sum_{i,j\in [n]} a_i b_i b_j 
	\parenv*{\parenv*{\sum{}_{k\in[n]} a_k} - (n-1) a_j - a_i} \\
	&= \sum_{i,j\in [n]} \sum_{k\in[n]\setminus\bracenv*{i}} 
	a_i b_i b_j (a_k - a_j) \\
	&= \sum_{i,j,k\in [n]} a_i b_i b_j (a_k - a_j) 
	- \sum_{i,j\in [n]} a_i b_i b_j (a_i - a_j) \\
	&= \frac{1}{2} \sum_{i,j,k\in [n]} a_i b_i (b_j - b_k) (a_k - a_j) \\
	&\phantom{=} - \frac{1}{2} \sum_{i,j\in [n]} b_i b_j (a_j - a_i)^2 
	\leq 0,
\end{align*}
in contradiction, where the last step follows from $(b_j - b_k) 
(a_k - a_j) \leq 0$ for all $j,k$.
\end{IEEEproof}

We can now prove the following bounds on the size of spheres:
\begin{theorem}\label{thm:sphere-lower-upper}
It holds that
$\binom{n}{r} \parenv*{\qmg-1}^r \leq 
s_r \leq \binom{n}{r} \parenv*{\qa-1}^r$.
\end{theorem}
\begin{IEEEproof}
For the right inequality, observe from 
\cref{thm:sphere-ratio-ave} that 
\begin{align*}
	s_r &= \prod_{k=0}^{r-1} \frac{s_{k+1}}{s_k} 
	\leq \prod_{k=0}^{r-1} \frac{(\qa-1) (n-k)}{k+1} 
	= \binom{n}{r} \parenv*{\qa-1}^r.
\end{align*}

On the other hand, from the arithmetic and geometric mean inequality 
we directly observe 
\begin{IEEEeqnarray*}{+rCl+x*}
    s_r 
    &=& \sum_{\substack{R\subseteq [n]\\ \abs*{R}=r}} \prod_{i\in R} (q_i-1) 
    \geq \binom{n}{r} \parenv[\Bigg]{\prod_{\substack{R\subseteq [n]\\ \abs*{R}=r}} 
    \prod_{i\in R} (q_i-1)}^{1\big/\binom{n}{r}} \\
    &=& \binom{n}{r} \parenv*{\prod_{i=1}^n (q_i-1)^{\binom{n-1}{r-1}}}^{1\big/\binom{n}{r}} 
    = \binom{n}{r} \parenv*{\qmg-1}^r.
	\\[-\normalbaselineskip] &&&\IEEEQEDhere
\end{IEEEeqnarray*}
\end{IEEEproof}

\begin{conjecture}
Observe that $s_1 = n (\qa-1)$ and $s_n = \parenv*{\qmg-1}^n$, 
achieving the upper and lower bounds of \cref{thm:sphere-lower-upper}, 
respectively. We conjecture that 
$\parenv*{s_r\big/\binom{n}{r}}^{1/r}$ is also decreasing, for $1\leq 
r\leq n$.
\end{conjecture}

\begin{corollary}\label{cor:ball-size-entropy}
For $0\leq r\leq \parenv*{1-\frac{1}{\qa}}n$ it holds that 
\begin{align*}
	\tfrac{1}{n+1} \qmg^{n H_{\qmg}(r/n)} \leq 
	\abs*{B_r(0)} \leq \qa^{n H_{\qa}(r/n)}.
\end{align*}
where $H_q(x) = x\log_q(q-1) - x\log_q(x) - (1-x)\log_q(1-x)$ is the 
$q$-ary entropy function.
\end{corollary}
\begin{IEEEproof}
We rely on the well-known bounds on the size of the $q$-ary Hamming 
ball (see, e.g., \cite[Lemmas 4.7-8]{Rot06}), and the bounds of 
\cref{thm:sphere-lower-upper}.
\end{IEEEproof}

\section{Johnson radius and list-decodability} \label{sec:list}

\begin{definition}
For $q>1$ and $\delta<1-\frac{1}{q}$ we denote the \emph{Johnson 
radius} 
\[
J_q(\delta)\deq \parenv*{1-\frac{1}{q}}\parenv*{1-\sqrt{1-\frac{\delta}{1-\frac{1}{q}}}}.
\]
\end{definition}
Observe that $\frac{\delta}{2}\leq J_q(\delta)\leq \delta < 
1-\frac{1}{q}$. Essentially, for $r < J_q(d/n)\cdot n$ it holds that 
$q r^2 > (q-1) (2r-d) n$.

We next follow the approach of~\cite[Lem.]{Bas65b}, and attributed to 
Johnson in~\cite[Th.~7.3.1]{GurRudSud19}, to bound $A_r(n,d), 
A_{\leq r}(n,d)$:
\begin{lemma} \label{lem:cnst-weight}
If $\qa r^2 > (\qa-1) (2r-d) n$ then 
\begin{align*}
    A_r(n,d) \leq A_{\leq r}(n,d) \leq 
    \frac{(\qa-1) n d}{\qa r^2 - (\qa-1) (2r-d) n}.
\end{align*}
\end{lemma}
\begin{IEEEproof}
That $A_r(n,d) \leq A_{\leq r}(n,d)$ follows from the definitions. 
Denote then $A\deq A_{\leq r}(n,d)$, and build a matrix with $A$ rows, 
each an element of a maximum-size code in $B_r(0)$. Hence 
\begin{enumerate}[label=(\alph*)]
    \item Every row has at most~$r$ nonzero coordinates. 
    (Here, the meaning of~$0$ depends on the column, but this fact has 
    no effect on our argument, and will be ignored).
    
    \item The coordinate-wise difference of any two distinct rows has 
    at least $d$ nonzero coordinates.
\end{enumerate}

For $1\leq i\leq n$ and any $j\in\bbZ/q_i\bbZ$, we let $k_{i,j}$ 
denote the number of incidences of $j$ in the $i$'s column of our 
matrix. Then we observe 
\begin{align*}
    \sum_{i=1}^n \sum_{0\neq j\in \bbZ/q_i\bbZ} k_{i,j} \leq A r; 
    \qquad \sum_{i=1}^n \sum_{j\in \bbZ/q_i\bbZ} k_{i,j} = A n.
\end{align*}
We note that the total number of nonzero elements in the $A(A-1)$ 
differences of any two distinct rows (when order is considered) is 
$\sum_{i=1}^n\sum_{j\in \bbZ/q_i\bbZ} k_{i,j} (A-k_{i,j})$. 
This number is at least $A(A-1) d$, by assumption. Then, denoting 
$\rho\deq \frac{1}{A}\sum_{i=1}^n \sum_{0\neq j\in \bbZ/q_i\bbZ} 
k_{i,j} \leq r$, 
\begin{IEEEeqnarray*}{+rCl+x*}
	\IEEEeqnarraymulticol{3}{l}{A(A-1) d 
	\leq \sum_{i=1}^n \sum_{j\in \bbZ/q_i\bbZ} k_{i,j} (A-k_{i,j})} \\
    \;\; &=& A \sum_{i=1}^n \sum_{j\in \bbZ/q_i\bbZ} k_{i,j} 
    - \sum_{i=1}^n \sum_{0\neq j\in \bbZ/q_i\bbZ} k_{i,j}^2 
    - \sum_{i=1}^n k_{i,0}^2 \\
    &\leq& A^2 n  - \frac{A^2 \rho^2}{\sum_{i=1}^n (q_i-1)} - 
    \frac{A^2 (n-\rho)^2}{n},
\end{IEEEeqnarray*}
where the last inequality uses Titu's lemma. By rearrangement of 
addends and multiplication by $\frac{(\qa-1) n}{A}$:
\begin{IEEEeqnarray*}{+rCl+x*}
	\IEEEeqnarraymulticol{3}{l}{(\qa-1) n d \geq 
	A \sparenv*{\qa\rho^2 - (\qa-1) (2\rho-d) n}} \\
    \;\; &=& A \sparenv*{(\qa-1) n d - \frac{(\qa-1)^2}{\qa} 
    \parenv*{n^2 - \parenv*{n - \frac{\rho}{1-1/\qa}}^2}} \\
    &\geq& A \sparenv*{(\qa-1) n d - \frac{(\qa-1)^2}{\qa} 
    \parenv*{n^2 - \parenv*{n - \frac{r}{1-1/\qa}}^2}} \\
    &=& A \sparenv*{\qa r^2 - (\qa-1) (2r-d) n}.
\end{IEEEeqnarray*}
The claim follows directly.
\end{IEEEproof}

\begin{corollary}
Take a code $C\subseteq \cQ$ with minimum distance~$d$, $0 < d < 
\parenv*{1-\frac{1}{\qa}} n$. 
For any $\rho\in\bbN$, $\rho < J_{\qa}(d/n)\cdot n$, and any 
$x\in\cQ$, it holds that $\abs*{C\cap B_\rho(x)}\leq (\qa-1) d n$.
\end{corollary}
\begin{IEEEproof}
From the shift-invariance of the Hamming distance, $C\cap B_\rho(x) = 
(C-x)\cap B_\rho(0)$, implying $\abs*{C\cap B_\rho(x)}\leq 
A_{\leq \rho}(n,d)$. Then, by assumption we have $\qa\rho^2 > (\qa-1) 
(2\rho-d) n$, and since all quantities are integers, $\qa\rho^2 \geq 
1 + (\qa-1) (2\rho-d) n$. Finally, observe from \cref{lem:cnst-weight} 
that $A_{\leq \rho}(n,d)\leq \frac{(\qa-1) n d}{\qa\rho^2 - (\qa-1) 
(2\rho-d) n} \leq (\qa-1) n d$, as required.
\end{IEEEproof}

The last corollary establishes a number of errors beyond 
$\floorenv*{\frac{d-1}{2}}$ in which codes with minimum distance~$d$ 
allow list-decoding with list size quadratic in~$n$, although unique 
decoding is no longer assured. Indeed, we note from $J_{\qa}(\delta) > 
\frac{\delta}{2}$ that for sufficiently large~$n$, $\rho\deq 
\ceilenv*{J_{\qa}(d/n)\cdot n}-1 > \floorenv*{\frac{d-1}{2}}$.

\section{Bounds} \label{sec:bounds}

In this section, we explore generalizations of known bounds on 
mono-alphabetic $q$-ary codes. We first present the known 
`Singleton-like' bound~\cite[Th.~2]{SidSchGabBosAfa05}, which is also 
developed as a `code-anticode' bound~\cite[Cor.~2.15]{Etz22} with the 
diameter-$(d-1)$ anticode $\parenv*{\prod_{i=1}^{n-d+1}\bracenv*{0}}
\times \parenv*{\prod_{i=n-d+2}^n (\Z/q_i\Z)}$:
\begin{theorem}\label{thm:known}
If $C\subseteq \cQ$ has minimum distance~$d$, then 
\begin{align*}
	\abs*{C}\leq \prod_{i=1}^{n-d+1} q_i.
\end{align*}
\end{theorem}

Next, we start with a corollary of the bounds of the last section; these are asymptotic version of the Gilbert-Varshamov and 
sphere-packing bounds.
\begin{corollary}\label{cor:asym-GV-SP}
For $0\leq \delta\leq 1-\frac{1}{\qa}$ it holds that 
\begin{multline*}
	1 - \tfrac{\log(\qa)}{\log(\qg)} H_{\qa}(\delta) + o(1) 
	\leq R(\delta) \\
	\leq 	1 - \tfrac{\log(\qmg)}{\log(\qg)} 
	H_{\qmg}(\tfrac{\delta}{2}) + o(1)
\end{multline*}
\end{corollary}
\begin{IEEEproof}
We utilize well-known proofs for these bounds (e.g., 
\cite[Ch.~4.5]{Rot06}), based on~\cref{cor:ball-size-entropy}.
\end{IEEEproof}

\subsection{Elias-Bassalygo bound}

In this section we pursue a parallel to the Elias-Bassalygo bound for 
mixed codes. We start with a proof of the \emph{Elias-Bassalygo 
inequality} (see, e.g., a corollary in \cite[Eq.~5]{Bas65b}), also 
referred to in~\cite[Eq.~2.2]{Etz22} as the \emph{local inequality 
lemma}.
\begin{lemma} \label{lem:pigeon}
For all $r\leq n$, it holds that $A(n,d)\leq \frac{\abs*{\cQ}}{s_r} 
A_r(n,d)$.
\end{lemma}
\begin{IEEEproof}
Take some maximum size code $C\subseteq\cQ$ with minimum distance $d$. 
We observe that for each $c\in C$, there exist exactly $s_r$ distinct 
$x\in\cQ$ such that $x+c\in S_r(0)$. It follows that $\sum_{x\in\cQ} 
\abs*{(x+C) \cap S_r(0)} = s_r \abs*{C} = s_r A(n,d)$. By the 
pigeonhole principle there must exist $x\in\cQ$ such that 
$\abs*{(x+C) \cap S_r(0)} \geq \frac{s_r A(n,d)}{\abs*{\cQ}}$. 
Note that $(x+C) \cap S_r(0)$ is a code in $S_r(0)$ with minimum 
distance $d$ (the Hamming distance is shift invariant), hence 
$A_r(n,d) \geq \frac{s_r A(n,d)}{\abs*{\cQ}}$, as required.
\end{IEEEproof}

\begin{corollary}\label{cor:bassalygo}
For $d<\parenv*{1-\frac{1}{\qa}} n$ and $r < J_{\qa}(d/n)\cdot n$ it 
holds that 
\begin{align*}
	A(n,d) &\leq \frac{\abs*{\cQ}}{s_r} \cdot 
	\frac{(\qa-1) n d}{\qa r^2 - (\qa-1) (2r-d) n}.
\end{align*}
\end{corollary}
\begin{IEEEproof}
The claim follows from \cref{lem:cnst-weight,lem:pigeon}.
\end{IEEEproof}

\begin{corollary}\label{cor:EB-asym}
For $\delta < \parenv*{1-\frac{1}{\qa}}$ it holds that 
\begin{align*}
	R(\delta) &\leq 1 - \tfrac{\log(\qmg)}{\log(\qg)}	 
	H_{\qmg}\parenv*{J_{\qa}(\delta)} + o(1).
\end{align*}
\end{corollary}
\begin{IEEEproof}
We denote $d_n\deq \floorenv*{\delta n}$ and let $r_n\deq \floorenv*{\parenv*{1-\frac{1}{\qa}}\sparenv*{1-\sqrt{1-\frac{(d_n-1) / n}{1-\frac{1}{\qa}}}} n}$; then by \cref{cor:bassalygo} 
$A(n,d_n)\leq \frac{\abs*{\cQ}}{s_{r_n}}\cdot d_n$.

Next, by \cref{thm:sphere-lower-upper} 
\begin{IEEEeqnarray*}{+rCl+x*}
	\log A(n,d_n) &\leq& \log\abs*{\cQ} - \log \tbinom{n}{r_n} - r_n \log(\qmg-1) + \log(d_n) \\
	&=& \log\abs*{\cQ} - \parenv*{\log(2) n H_2(r_n/n) - O(\log n)} \\
	&&-\> r_n \log(\qmg-1) + \log(d_n) \\
	&=& \log\abs*{\cQ} - \log(\qmg) n H_{\qmg}(r_n/n) + O(\log n),
\end{IEEEeqnarray*}
which concludes the proof.
\end{IEEEproof}
Before concluding, we observe from $J_{\qa}(\delta)>\frac{\delta}{2}$ 
that \cref{cor:EB-asym} is tighter than the sphere-packing upper bound 
of \cref{cor:asym-GV-SP}.

\subsection{First linear-programming bound}

We adapt techniques utilized in \cite{BhaBan18} for a Fourier analytic 
approach to the development of the first linear-programming bound on 
$q$-ary codes (first developed in the binary case by McEliece et al. 
\cite{McERodRumWel77} and later generalized by Levenshtein 
\cite{Lev98}), to the case of mixed codes. These methods draw on a 
similar treatment of the binary case in \cite{NavSam09}. We stress 
that \cite{BhaBan18} proved the bound in the mono-alphabetic $q$-ary 
case for a class of \emph{symmetric} codes (containing all linear 
codes, and, in the binary case, all codes); our analysis shows that 
this requirement can be dropped (since $q$-ary codes are a private 
case of our setting, our results directly apply to that case as well). 
We also choose to more rigorously treat the Fourier duality.

The Hamming metric on $\cQ$ is fully represented by the graph on vertex set $\cQ$, where $x,y\in\cQ$ are connected by an (undirected) edge if and only if $d(x,y)=1$. That is, for any $x,y\in\cQ$ it holds that $d(x,y)$ is also the \emph{graph distance} of $x,y$ (i.e., the length of the shortest path in the graph between $x,y$). Motivated by this fact, throughout this section we let $\cA$ be the \emph{adjacency matrix} of this graph. By abuse of notation, we let $\cA$ operate on functions $f\colon \cQ\to\bbC$ by considering $f$ to be an $\abs*{\cQ}$-tuple over $\bbC$ (indexed identically to $\cA$).

The purpose of this section is to develop an upper bound on 
$R(\delta)$, parallel to the first linear-programming bound. The 
methods we use rely on Fourier analysis in finite Abelian groups, 
therefore in the interest of self-containment and clear notation we 
dedicate \cref{sec:fourier} to review basic notions, mostly based 
on~\cite{Con10}. \Cref{sec:bound-by-ev,sec:ball-ev} are dedicated to 
prove the two main propositions necessary for this section, 
\cref{th:bound-by-ev,cor:ball-eigenvalue}.

\begin{definition}
For a subset $B\subseteq\cQ$, we define the \emph{maximum eigenvalue of $B$} by 
\[
\lB\deq \max\mathset*{\frac{\inner{\cA f}{f}}{\inner{f}{f}}}{f\colon\cQ\to\bbR,\ \supp(f)\subseteq B}.
\]
It is the maximum eigenvalue of the minor of $\cA$ corresponding to $B$; i.e., the adjacency matrix of the subgraph of the Hamming graph spanned by the elements of $B$. Since the entries of $\cA$ are non-negative, by Perron's theorem, $\lB$ is non-negative, greater than or equal to the absolute value of any other eigenvalue, and there exists a non-negative eigenfunction $\fB\colon\cQ\to\bbR$, $\fB\geq 0$, of $\lB$ (that is, $\cA \fB = \lB \fB$).
\end{definition}

\begin{theorem}\label{th:bound-by-ev}
Let $C\subseteq\cQ$ be a code with minimum distance $d$. Take a 
symmetric $B\subseteq\cQ$ such that $\lB\geq (n+1) (\qa-1) - 
\sum_{i=1}^d q_i$. Then 
\begin{align*}
    \abs*{C}\leq n\abs*{B}.
\end{align*}
\end{theorem}

\begin{corollary}\label{cor:ball-eigenvalue}
Take $2\sqrt{n}<r\leq n$. Then $\lambda_{B_r(0)} \geq 
2\sqrt{(\qa-1) r (n-r)} + (\qa-2) r + o(n)$.
\end{corollary}

Based on these results, we can now show the following.
\begin{theorem}\label{th:lp-asym}
Denote 
\begin{align*}
	\rho = \frac{n}{\qa} 
	\parenv*{(\qa-1) - (\qa-2) \check{\delta} 
	- 2 \sqrt{(\qa-1) \check{\delta} \parenv*{1 - \check{\delta}}}},
\end{align*}
where $\check{\delta}\deq \frac{1}{\qa n} \sum_{i=1}^d q_i$. 
If $C\subseteq \cQ$ has minimum distance~$d$, then 
\begin{align*}
	\log\abs*{C} \leq n\log(\qa) H_{\qa}(\rho/n) + o(n).
\end{align*}
\end{theorem}
\begin{IEEEproof}
%
We find the solution in $r$ to the equation $2\sqrt{(\qa-1) r (n-r)} + 
(\qa-2) r = (\qa-1) n - \sum_{i=1}^d q_i = (\qa-1) n - \qa n 
\check{\delta}$. Indeed, this equation holds 
if and only if 
\begin{align*}
	4(\qa-1) r (n-r) &= (\qa-2)^2 r^2 \\
	&\phantom{=}- 2(\qa-2) \parenv*{(\qa-1) n - \qa n \check{\delta}} r \\
	&\phantom{=}+ \parenv*{(\qa-1) n - \qa n \check{\delta}}^2 \\
	\iff 
	0 &= \qa^2 r^2 \\
	&\phantom{=}- 2n\qa \parenv*{(\qa-1) - (\qa-2) \check{\delta}} r \\
	&\phantom{=}+ n^2 \parenv*{(\qa-1) - \qa \check{\delta}}^2.
\end{align*}
This implies the two solutions 
\begin{align*}
	r_{\pm} = \frac{n}{\qa} 
	\parenv*{(\qa-1) - (\qa-2) \check{\delta} 
	\pm 2 \sqrt{(\qa-1) \check{\delta} \parenv*{1 - \check{\delta}}}}
\end{align*}
(trivially, $\rho=r_-<r_+$). 
It follows from \cref{th:bound-by-ev,cor:ball-eigenvalue} that 
$\abs*{C}\leq n \abs*{B_r(0)}$ for $r = \rho + o(n)$, and from 
\cref{cor:ball-size-entropy} we obtain the claim.
\end{IEEEproof}

\section{Conclusion}\label{sec:conc}

In conclusion, we plot the lower bound of \cref{cor:asym-GV-SP} with 
the upper bounds of \cref{cor:EB-asym,th:lp-asym}, in comparison to 
the known upper bound of \cref{thm:known}, in some special cases. 
Hence, we demonstrate that our proven bounds compete with the known 
one for some choices of alphabet sizes and normalized minimum 
distance. In principle, our bounds are more competitive for smaller 
alphabet sizes (which can be expected, given that the Singleton bound 
is already competitive in the mono-alphabetic $64$-ary case), and for 
distributions of alphabet sizes closer to the mono-alphabetic case.

A natural question is whether efficiently en-/decodable codes can be 
constructed to approach these bounds; we delegate an answer for this 
question to a future work.

\begin{figure}[t]
	\centering
	\includegraphics[width=\columnwidth]{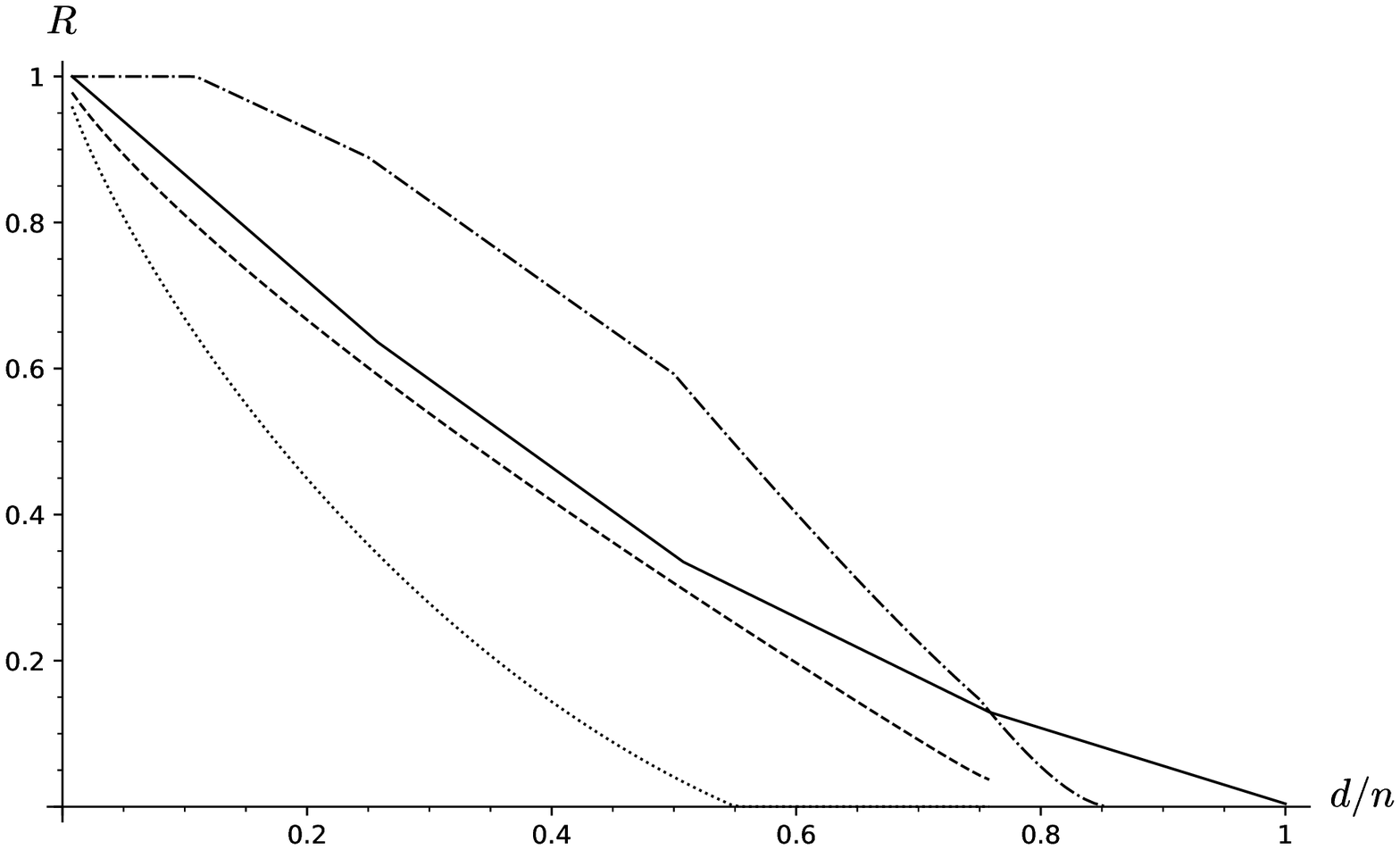}
	\caption{GV (\cref{cor:asym-GV-SP}, \texttt{$\cdot\cdot$}); EB (\cref{cor:EB-asym}, \texttt{-$\cdot$}); LP (\cref{th:lp-asym}, \texttt{-\;\!-}) and S (\cref{thm:known}, \texttt{-}) bounds. Alphabet sizes $[2,3,5,7]$ ($25\%,25\%,25\%,25\%$)\label{fig:q2-3-5-7_inc25-25-25-25}}
\end{figure}

\begin{figure}[t]
	\centering
	\includegraphics[width=\columnwidth]{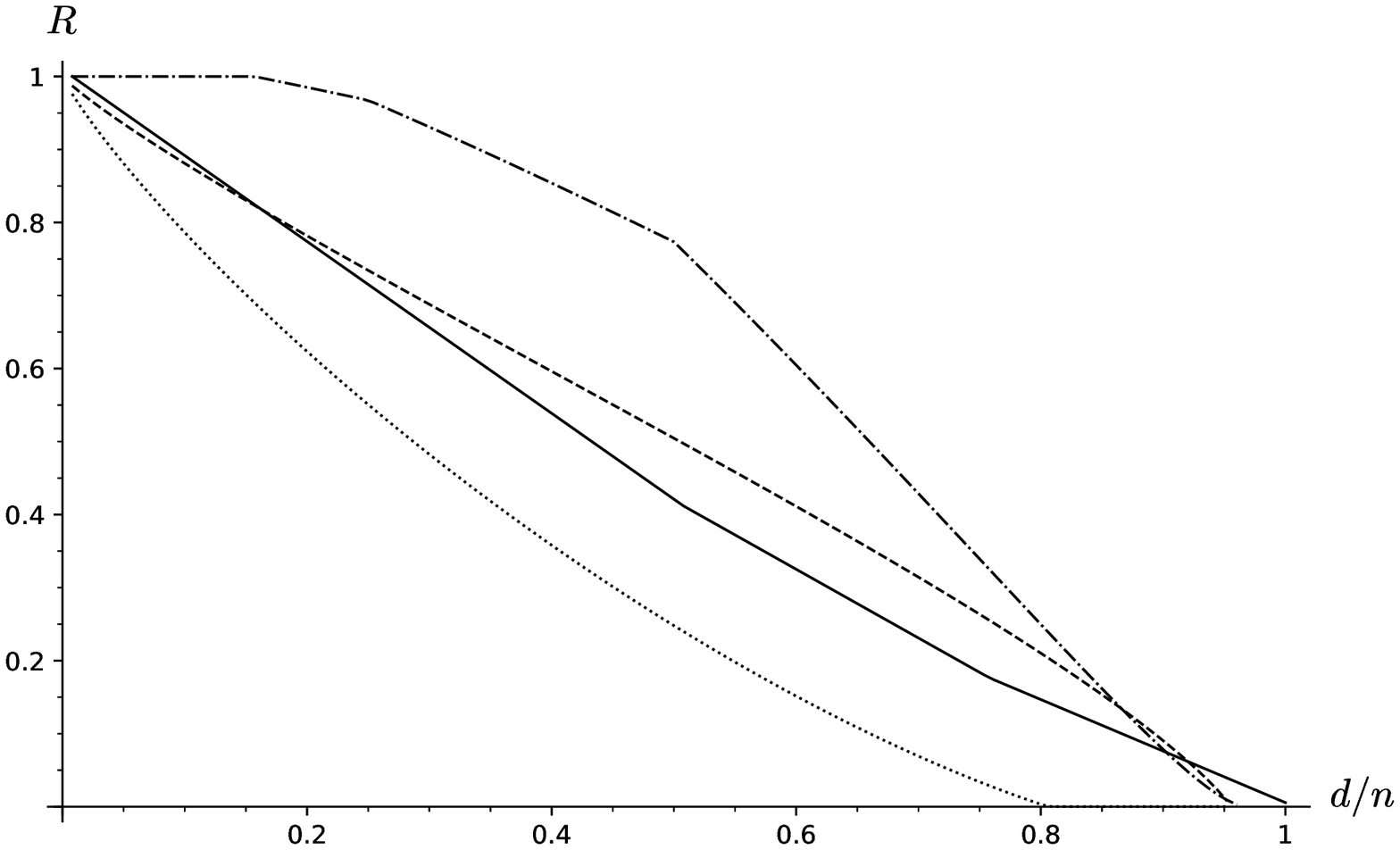}
	\caption{GV (\cref{cor:asym-GV-SP}, \texttt{$\cdot\cdot$}); EB (\cref{cor:EB-asym}, \texttt{-$\cdot$}); LP (\cref{th:lp-asym}, \texttt{-\;\!-}) and S (\cref{thm:known}, \texttt{-}) bounds. Alphabet sizes $[8,16,32]$ ($25\%,25\%,50\%$)\label{fig:q8-16-32_inc25-25-50}}
\end{figure}

\begin{figure}[t]
	\centering
	\includegraphics[width=\columnwidth]{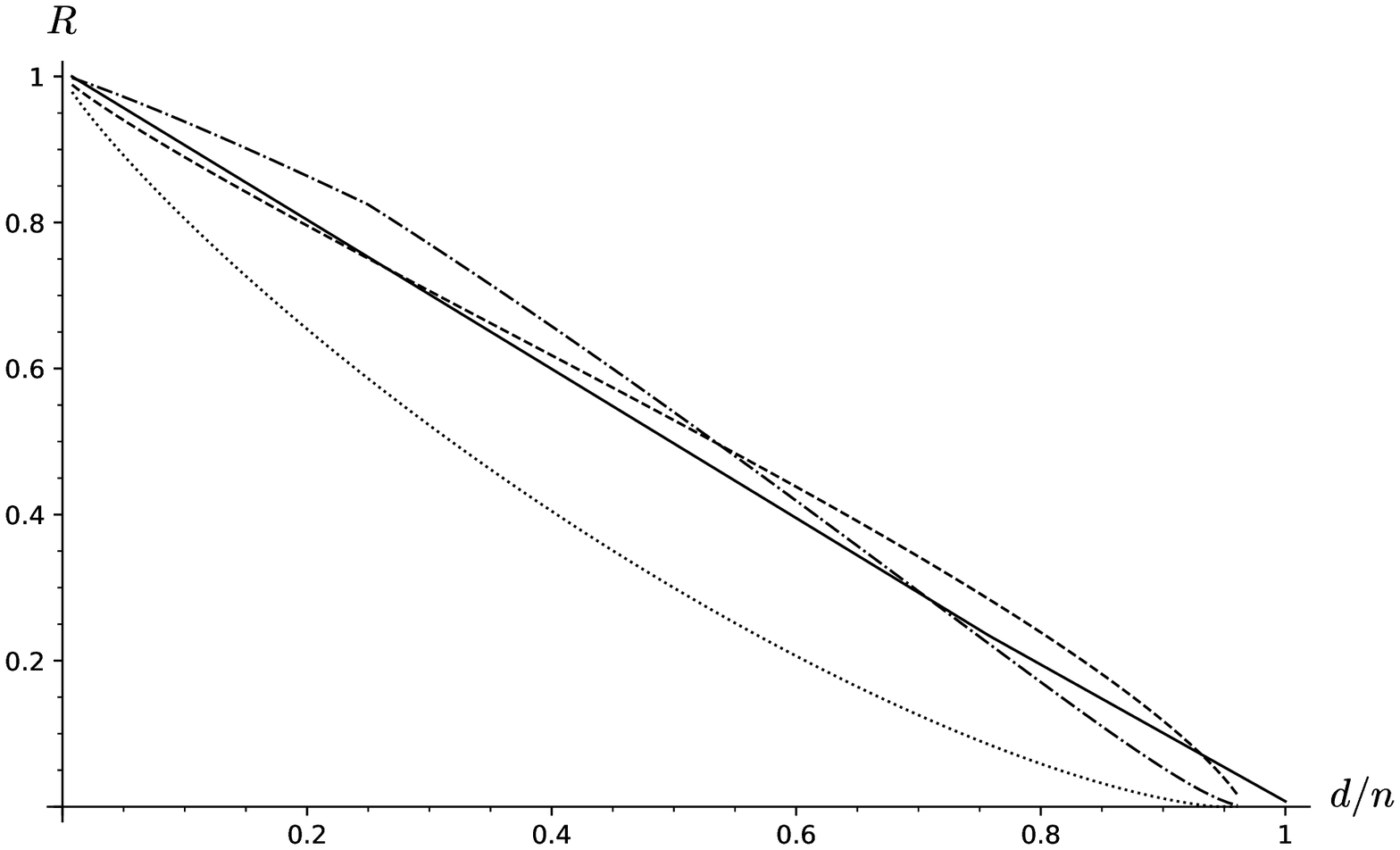}
	\caption{GV (\cref{cor:asym-GV-SP}, \texttt{$\cdot\cdot$}); EB (\cref{cor:EB-asym}, \texttt{-$\cdot$}); LP (\cref{th:lp-asym}, \texttt{-\;\!-}) and S (\cref{thm:known}, \texttt{-}) bounds. Alphabet sizes $[24,32]$ ($25\%,75\%$)\label{fig:q24-32_inc25-75}}
\end{figure}

\begin{figure}[t]
	\centering
	\includegraphics[width=\columnwidth]{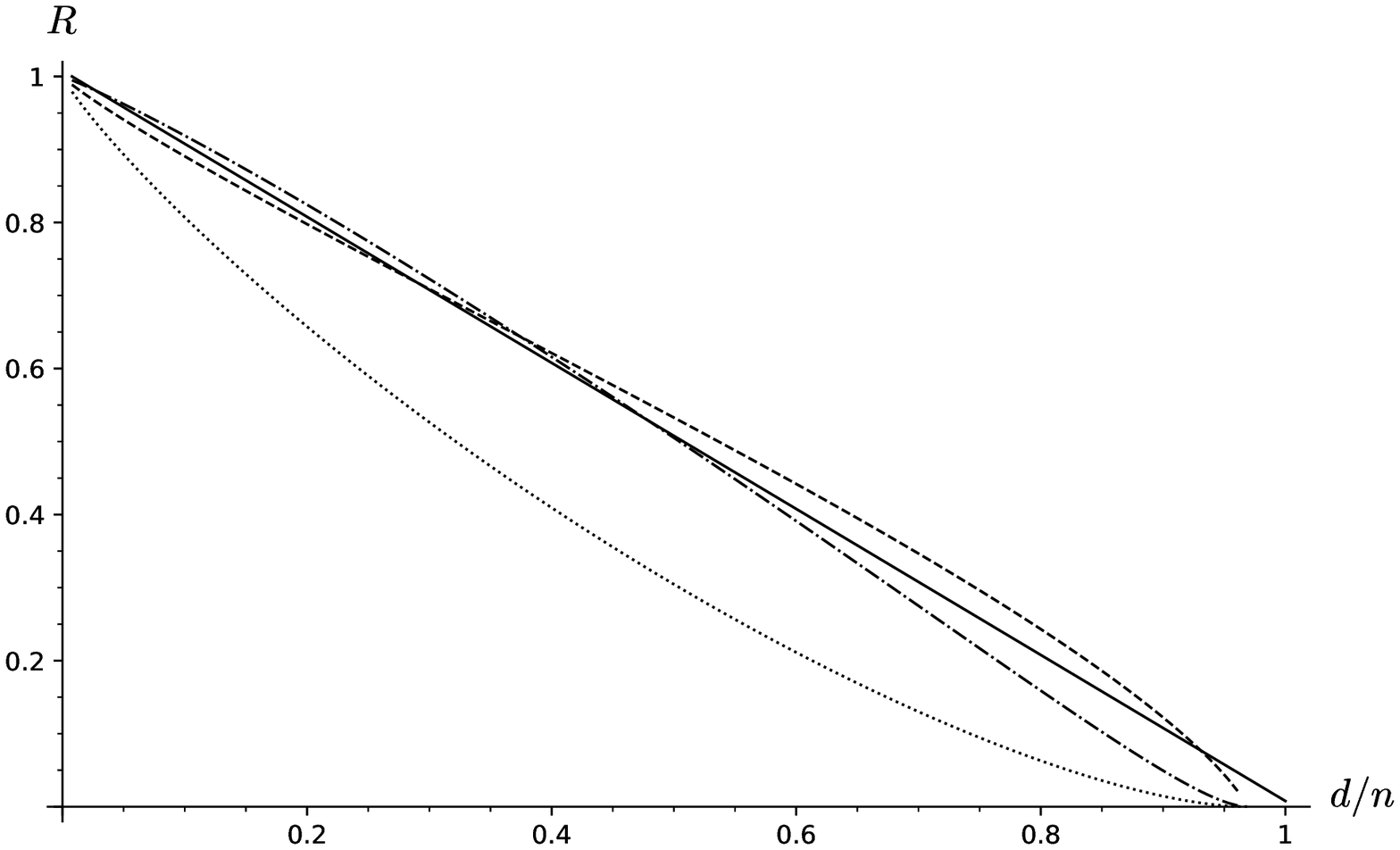}
	\caption{GV (\cref{cor:asym-GV-SP}, \texttt{$\cdot\cdot$}); EB (\cref{cor:EB-asym}, \texttt{-$\cdot$}); LP (\cref{th:lp-asym}, \texttt{-\;\!-}) and S (\cref{thm:known}, \texttt{-}) bounds. Alphabet sizes $[32]$ ($100\%$)\label{fig:q32_inc1}}
\end{figure}

\appendix

\subsection{Fourier Analysis of finite Abelian groups}\label{sec:fourier}

Let $G$ be a finite Abelian group (we shall use `$+$' to denote its group action, and `$0$' to denote its identity). A \emph{character} of $G$ is a group homomorphism $\chi\colon G\to S^1\subseteq \bbC$ (where the latter is equipped with complex multiplication as group action). The set~$\widehat{G}$ of all characters of $G$, endowed with the action of pointwise multiplication $(\chi\cdot\psi)(g)\deq (\chi(g))(\psi(g))$, is also an Abelian group (owing to the properties of multiplication of complex numbers), with $\chi^{-1} = \overline{\chi}$ (here, $\overline{\chi}(g)\deq \overline{\chi(g)}$). Its identity is the constant character $\1(g)\equiv 1$.

\begin{lemma}
Let $\chi$ be a character of a finite Abelian group $G$. For all $g\in G$ it holds that $\overline{\chi(g)} = \chi(-g)$
\end{lemma}
\begin{IEEEproof}
We observe $(\chi(g)) (\overline{\chi(g)}) = (\chi\cdot \chi^{-1})(g) = 1 = \chi(0) = \chi(g-g) = (\chi(g)) (\chi(-g))$. The claim follows.
\end{IEEEproof}

\begin{lemma}\cite[Thm.~4.1]{Con10}\label{lem:sum-of-char}
Let $\chi$ be a character of a finite Abelian group $G$. Then 
\[
\sum_{g\in G} \chi(g) = \begin{cases}
\abs*{G}, & \chi = \1; \\
0, & \text{otherwise}.
\end{cases}
\]
Dually, for any $g\in G$ 
\[
\sum_{\chi\in \widehat{G}} \chi(g) = \begin{cases}
\abs*{G}, & g = 0; \\
0, & \text{otherwise}.
\end{cases}
\]
\end{lemma}


\begin{definition}\label{def:char-prod}
For a given representation of $G$ as a direct product of finite Abelian groups $G=\prod_{i=1}^n G_i$, we make the following definitions: 
\begin{itemize}
\item The \emph{weight} of $g = (g_1,\ldots,g_n)\in G$ is defined $\wt(g)\deq \abs*{\mathset*{i\in[n]}{g_i\neq \1_{G_i}}}$.

\item For $\chi\in\widehat{G}$, $\chi_i\in\widehat{(G_i)}$ is defined for $g_i\in G_i$: $\chi_i(g_i)\deq \chi\parenv*{\1_{G_1},\ldots,\1_{G_{i-1}},g_i,\1_{G_{i+1}},\ldots,\1_{G_n}}$. It is straightforward to verify that $\chi_i$ is indeed a character of $G_i$, that $\chi(g_1,\ldots,g_n) = \prod_{i=1}^n \chi_i(g_i)$, and that this is a group homomorphism $\widehat{G}\to\prod_{i=1}^n \widehat{G_i}$.

\item For $\chi\in\widehat{G}$, its \emph{weight} is defined $\wt(\chi)\deq \abs*{\mathset*{i\in[n]}{\chi_i\neq \1_{\widehat{G_i}}}}$.
\end{itemize}
\end{definition}

The following propositions appeared in \cite{Con10} in very similar form; in the sequel, their locations are indicated, and the proofs are reproduced to include new observations.
\begin{lemma}\cite[Thm.~3.11]{Con10}\label{lem:cyc-dual}
If $G$ is cyclic then $G\cong\widehat{G}$ as groups; one such isomorphism, denoted $g\mapsto\widehat{g}$, satisfies for all $g_1,g_2\in G$: $\widehat{g_1}(g_2) = \widehat{g_2}(g_1)$.
\end{lemma}
\begin{IEEEproof}
First, we show that $\widehat{G}$ is cyclic, and since by \cite[Thm.~3.5]{Con10} $n\deq \abs[\big]{\widehat{G}} = \abs*{G}$, they are isomorphic.

Take a generator $\bar{g}$ of $G$, and set $\chi(\bar{g}^t) \deq e^{2\pi i t/n}$ (then, indeed, $\chi\in\widehat{G}$). For any $\psi\in\widehat{G}$, observe that $\psi(\bar{g})^n=\psi(\bar{g}^n)=\psi(\1_G)=1$, hence there exists an integer $k$ such that $\psi(\bar{g}) = e^{2\pi i k/n}$. Then, for all $t$ it holds that $\psi(\bar{g}^t) = \psi(\bar{g})^t = e^{2\pi i k t/n} = \chi(\bar{g}^t)^k = (\chi^k)(\bar{g}^t)$, hence $\psi=\chi^k$, as required.

We denote, then, $\widehat{\bar{g}^t} \deq \chi^t$, a group isomorphism. To conclude the proof, observe that $\widehat{\bar{g}^t}(\bar{g}^s) = \chi^t(\bar{g}^s) = \chi^{ts}(\bar{g}) = \chi^s(\bar{g}^t) = \widehat{\bar{g}^s}(\bar{g}^t)$.
\end{IEEEproof}

It is interesting to note that the isomorphism in the proof of the last lemma is not `canonical' or unique. Indeed, any distinct choice of generator $\bar{g}$ yields a distinct isomorphism. All, however, satisfy the last statement of the lemma.

\begin{lemma}\cite[Lem.~3.12]{Con10}\label{lem:prod-dual}
If $A,B$ are finite Abelian groups, then $\widehat{A\times B}, \widehat{A}\times\widehat{B}$ are isomorphic as groups.
\end{lemma}
\begin{IEEEproof}
Consider the homomorphism $\widehat{A\times B}\to \widehat{A}\times\widehat{B}$ described in \cref{def:char-prod}. Observe that its kernel is trivial, and that $\abs[\big]{\widehat{A\times B}} = \abs*{A}\cdot\abs*{B} = \abs[\big]{\widehat{A}\times\widehat{B}}$. Hence, it is a group isomorphism.
\end{IEEEproof}

\begin{corollary}\label{cor:prod-dual-inverse}
Let $A,B$ be finite Abelian groups. Given isomorphisms of $A,\widehat{A}$ and $B,\widehat{B}$ satisfying for all $a_1,a_2\in A$, $b_1,b_2\in B$ that $\widehat{a_1}(a_2)=\widehat{a_2}(a_1)$ and $\widehat{b_1}(b_2)=\widehat{b_2}(b_1)$, there is an isomorphism of $A\times B, \widehat{A\times B}$ satisfying $\widehat{(a_1,b_1)}(a_2,b_2) = \widehat{(a_2,b_2)}(a_1,b_1)$.
\end{corollary}
\begin{IEEEproof}
Note for $a\in A$, $b\in B$ that $(a,b)\mapsto (\widehat{a},\widehat{b})$ is a group isomorphism $A\times B\to \widehat{A}\times\widehat{B}$. 
From the last lemma, we may complete it with the isomorphism $\widehat{A}\times\widehat{B}\to\widehat{A\times B}$ defined for $a,a'\in A$ and $b,b'\in B$ by $(\widehat{a},\widehat{b})(a',b') = \widehat{a}(a') \widehat{b}(b')$. 
Denote the composition of both isomorphisms $(a,b)\mapsto \widehat{(a,b)}$. 
Finally, observe that $\widehat{(a_1,b_1)}(a_2,b_2) = \widehat{a_1}(a_2) \widehat{b_1}(b_2) = \widehat{a_2}(a_1) \widehat{b_2}(b_1) = \widehat{(a_2,b_2)}(a_1,b_1)$.
\end{IEEEproof}

\begin{theorem}\cite[Thm.~3.13]{Con10}\label{thm:dual}
For any finite Abelian group $G$, $G\cong\widehat{G}$ as groups; one such isomorphism, denoted $g\mapsto\widehat{g}$, satisfies for all $g_1,g_2\in G$: $\widehat{g_1}(g_2) = \widehat{g_2}(g_1)$.

Furthermore, given a decomposition of $G$ as a direct product of finite \emph{cyclic} groups $G=\prod_{i=1}^n G_i$, it holds for all $g\in G$ that $\wt(\widehat{g}) = \wt(g)$.
\end{theorem}
\begin{IEEEproof}
The claim was proven in \cref{lem:cyc-dual} for the case that $G$ is cyclic. Recall that every finite Abelian group $G$ is isomorphic to a product of cyclic groups; then, \cref{lem:cyc-dual} and \cref{cor:prod-dual-inverse} readily extend to a product $G=\prod_{i=1}^n G_i$.

For the last part, observe that regardless of choice of generator in the proof of \cref{lem:cyc-dual}, $\widehat{\1_{G_i}} = \1_{\widehat{G_i}}$.
\end{IEEEproof}

Next, we observe some properties of complex functions of a finite Abelian group.
\begin{definition}[Expected value and inner product]
We define the \emph{expected value} of a function $f\colon G\to\bbC$ by 
\begin{align}
    \bbE(f)\deq \frac{1}{\abs*{G}}\sum_{g\in G} f(g).
\end{align}
This allows us to also define an inner product of functions $f,h\colon G\to\bbC$ by 
\begin{align}
    \inner{f}{h} \deq \bbE(f\overline{h}) = \frac{1}{\abs*{G}}\sum_{g\in G} f(g)\overline{h(g)}.
\end{align}
(This is in fact the normalized standard inner product, when we identify $f\colon G\to\bbC$ with $f\in \bbC^{\abs*{G}}$.)
Likewise, we define the expected value of a function $F\colon \widehat{G}\to\bbC$ by 
\begin{align}
    \bbE(F)\deq \frac{1}{\abs*{G}}\sum_{\chi\in\widehat{G}} F(\chi).
\end{align}
(Recall by \cite[Thm.~3.5]{Con10} that $\abs{\widehat{G}}=\abs*{G}$.)

In contrast to function on $G$, we define the inner product of two functions $F,H\colon\widehat{G}\to\bbC$ by 
\begin{align}
    \inner{F}{H} \deq \sum_{\chi\in\widehat{G}} F(\chi)\overline{H(\chi)}.
\end{align}
\end{definition}

\begin{corollary}\label{cor:ortho-char}\cite[Th.~A.2]{BhaBan18}
Let $\chi,\psi$ be any two characters of a finite Abelian group $G$. Then 
\[
\inner{\chi}{\psi} = \begin{cases}
1, & \chi = \psi; \\
0, & \text{otherwise}.
\end{cases}
\]
\end{corollary}

\begin{definition}
For a function $f\colon G\to\bbC$, its \emph{Fourier transform} $\widehat{f}\colon \widehat{G}\to \bbC$ is defined by 
\begin{align}
    \widehat{f}(\chi)\deq \inner{f}{\chi} = \frac{1}{\abs*{G}} \sum_{g\in G} f(g) \overline{\chi(g)}.
\end{align}
\end{definition}

\begin{lemma}\label{lem:fourier-sum}
For a function $f\colon G\to\bbC$ it holds that 
\[
f = \sum_{\chi\in\widehat{G}} \widehat{f}(\chi) \chi.
\]
\end{lemma}
\begin{IEEEproof}
We observe for all $g\in G$ that 
\begin{align}
    \sum_{\chi\in\widehat{G}} \widehat{f}(\chi) \chi(g) 
    &= \sum_{\chi\in\widehat{G}} \parenv*{\frac{1}{\abs*{G}} \sum_{g'\in G} f(g') \overline{\chi(g')}} \chi(g) \\
    &= \sum_{g'\in G} f(g') \parenv*{\frac{1}{\abs*{G}} \sum_{\chi\in\widehat{G}} \chi(g) \overline{\chi(g')}} \nonumber\\
    &= \sum_{g'\in G} f(g') \parenv*{\frac{1}{\abs*{G}} \sum_{\chi\in\widehat{G}} \chi(g-g')} 
    = f(g),
\end{align}
where the last step is justified by \cref{lem:sum-of-char}. This concludes the proof.
\end{IEEEproof}

\begin{lemma}\label{lem:even-real}
Take $f\colon G\to\bbC$, and let $f^-(g)\deq f(-g)$, $f^*(g)\deq 
\overline{f(g)}$. Then 
\begin{enumerate}
\item
$\widehat{f^-}(\chi) = \widehat{f}(\chi^{-1})$. In particular, $f$ is 
even (i.e., $f(-g)=f(g)$ for all $g\in G$) if and only if 
$\widehat{f}$ is even (i.e., $\widehat{f}(\chi^{-1}) = 
\widehat{f}(\chi)$ for all $\chi\in\widehat{G}$).

\item
$\widehat{f^*}(\chi) = \overline{\widehat{f}(\chi^{-1})}$. In 
particular, if $f,\widehat{f}$ are even, then $f$ is real-valued if 
and only if $\widehat{f}$ is real-valued.
\end{enumerate}
\end{lemma}
\begin{IEEEproof}
For the first part, note that 
\begin{align*}
	\widehat{f^-}(\chi) 
	&= \frac{1}{\abs*{G}} \sum_{g\in G} f^-(g) \overline{\chi(g)} 
	= \frac{1}{\abs*{G}} \sum_{g\in G} f(-g) \chi(-g) \\
	&= \frac{1}{\abs*{G}} \sum_{g'\in G} f(g') \overline{\chi^{-1}(g')} 
	= \widehat{f}(\chi^{-1}).
\end{align*}
(Recall due to \cref{lem:fourier-sum} that the Fourier transform is 
injective.)

For the second part, we note 
\begin{align*}
	\widehat{f^*}(\chi) 
	&= \frac{1}{\abs*{G}} \sum_{g\in G} f^*(g) \overline{\chi(g)} 
	= \overline{\frac{1}{\abs*{G}} \sum_{g\in G} f(g) \chi(g)} \\
	&= \overline{\frac{1}{\abs*{G}} \sum_{g\in G} f(g) 
	\overline{\chi^{-1}(g)}} 
	= \overline{\widehat{f}(\chi^{-1})}.
\end{align*}
\end{IEEEproof}

We also have the following familiar result:
\begin{theorem}[Plancharel's Theorem]\label{thm:Plancharel}\cite[Th.~A.3]{BhaBan18}
For two functions $f,h\colon G\to\bbC$ it holds that 
\[
\inner{f}{h} = \inner{\widehat{f}}{\widehat{h}}.
\]
\end{theorem}

\begin{definition}[Convolution]
For an element $g\in G$ and a function $f\colon G\to\bbC$, we let $g$ act on $f$ by $f^g(g')\deq f(g-g')$.

The \emph{convolution} of two functions $f,h\colon G\to\bbC$ is defined by 
\begin{align*}
	f*h(g)\deq \bbE(f h^g) 
	&= \frac{1}{\abs*{G}} \sum_{g'\in G} f(g') h^g(g') \\
	&= \frac{1}{\abs*{G}} \sum_{g'\in G} f(g') h(g-g').
\end{align*}
For two functions $F,H\colon\widehat{G}\to\bbC$, we define their \emph{convolution} by 
\[
F*H(\chi)\deq \sum_{\psi\in\widehat{G}} F(\psi) H(\chi\cdot\psi^{-1}).
\]
\end{definition}
It can be verified that both convolution operators are associative and commutative.

\begin{lemma}\label{lem:conv-even}
Let $f\colon G\to\bbC$ be even. Then $f*f$ is also even.
\end{lemma}
\begin{IEEEproof}
For all $g\in G$ 
\begin{align*}
    f * f (-g) &= \bbE(f f^{-g}) 
    = \frac{1}{\abs*{G}} \sum_{g'\in G} f(g') f(-g-g') \\
    &= \frac{1}{\abs*{G}} \sum_{g'\in G} f(g') f(g+g') \\
    &= \frac{1}{\abs*{G}} \sum_{g''\in G} f(g''-g) f(g'') \\
    &= \frac{1}{\abs*{G}} \sum_{g''\in G} f(g-g'') f(g'') 
    = f * f (g).
\end{align*}
\end{IEEEproof}

In familiar fashion, the Fourier transform translates between convolution and point-wise multiplication:
\begin{theorem}\cite[Th.~A.5]{BhaBan18}\label{thm:conv-to-prod}
For a character~$\chi$ of a finite Abelian group~$G$, and two functions $f,h\colon G\to\bbC$, 
\[
\widehat{f*h}(\chi) = \widehat{f}(\chi) \widehat{h}(\chi).
\]
\end{theorem}
\begin{theorem}\label{thm:prod-to-conv}
For a character~$\chi$ of a finite Abelian group~$G$, and two functions $f,h\colon G\to\bbC$, 
\[
\widehat{f h}(\chi) = \widehat{f}*\widehat{h}(\chi).
\]
\end{theorem}
\begin{IEEEproof}
\begin{align*}
    \widehat{f}*\widehat{h}(\chi) 
    &= \sum_{\psi\in\widehat{G}} \widehat{f}(\psi) \widehat{h}(\chi\cdot\psi^{-1}) \\
    &= \frac{1}{\abs*{G}^2} \sum_{\psi\in\widehat{G}} \sum_{g,g'\in G} 
    f(g) \overline{\psi(g)} h(g') \overline{\chi(g')} \psi(g') \\
    &= \frac{1}{\abs*{G}^2} \sum_{g,g'\in G} f(g) h(g') \overline{\chi(g')} 
    \sum_{\psi\in\widehat{G}} \psi(g'-g) \\
    &= \frac{1}{\abs*{G}} \sum_{g\in G} f(g) h(g) \overline{\chi(g)} 
    = \widehat{f h}(\chi).
\end{align*}
\end{IEEEproof}

The following lemma will be useful in our subsequent analysis:
\begin{lemma}[Interchange Lemma]\cite[Lem.~A.7]{BhaBan18}
Let $f_{1,2,3,4}\colon G\to\bbC$ be any four functions, such that the \emph{Fourier coefficients} $\bracenv*{\widehat{f_i}(\chi)}\big._{\chi\in\widehat{G}}$ are all real for $i=1,2,3,4$. It holds that 
\[
\inner{f_1*f_2}{f_3*f_4} = \inner{f_2}{f_1*f_3*f_4} = \inner{f_3*f_1*f_2}{f_4}.
\]
\end{lemma}

Finally, we consider another mapping from functions on $G$ to functions on $\widehat{G}$:
\begin{definition}
For $f\colon G\to\bbC$, define its \emph{dual function} $\widecheck{f}\colon\widehat{G}\to\bbC$ by $\widecheck{f}(\widehat{g})\deq f(g)$.
\end{definition}

It is natural to consider the dual to \cref{lem:fourier-sum}:
\begin{lemma}\label{lem:dual-transform}
For a function $f\colon G\to\bbC$, we may define $f'\colon G\to\bbC$ by 
\[
f'(g)\deq \sum_{\chi\in\widehat{G}} \widecheck{f}(\chi) \chi(g).
\]
(That is, $\widehat{f'} = \widecheck{f}$.) 
Then, it holds for all $g\in G$ that $f'(g) = \abs*{G} \widehat{f}(\widehat{g}^{-1})$.
\end{lemma}
\begin{IEEEproof}
Indeed, 
\begin{align*}
    f'(g) &= \sum_{\chi\in\widehat{G}} \widecheck{f}(\chi) \chi(g) = \sum_{g'\in G} \widecheck{f}(\widehat{g'}) \widehat{g'}(g) 
    = \sum_{g'\in G} f(g') \widehat{g'}(g) \\
    &= \sum_{g'\in G} \parenv[\bigg]{\sum_{\chi\in\widehat{G}} \widehat{f}(\chi) \chi(g')} \widehat{g'}(g) \\
    &= \sum_{\chi\in\widehat{G}} \widehat{f}(\chi) \parenv[\bigg]{\sum_{g'\in G} \chi(g') \widehat{g'}(g)} \\
    &= \sum_{\chi\in\widehat{G}} \widehat{f}(\chi) \parenv[\bigg]{\sum_{g'\in G} (\chi\cdot\widehat{g})(g')} 
    = \abs*{G} \widehat{f}(\widehat{g}^{-1}).
\end{align*}
\end{IEEEproof}

\subsection{Upper bound using subset eigenvalue}\label{sec:bound-by-ev}

\begin{definition}
Define a function $L\colon\cQ\to\bbR$ by 
\[
L(x) \deq \begin{cases}
\abs*{\cQ}, & \wt(x)=1; \\
0, & \text{otherwise}.
\end{cases}
\]
\end{definition}

This definition is useful since it offers an alternative representation of $\cA f$ for any $f\colon\cQ\to\bbC$ as follows:
\begin{lemma}
If $f\colon\cQ\to\bbC$ then $\cA f = f * L$.
\end{lemma}
\begin{IEEEproof}
Observe, for all $x\in\cQ$, 
\begin{align*}
    f * L(x) &= \bbE(f L^x) 
    = \sum_{y\in \cQ} f(y) \frac{L(x-y)}{\abs*{\cQ}} 
    = \sum_{\substack{y\in\cQ \\ d(x,y)=1}} f(y) \\
    &= \sum_{y\in \cQ} \cA(x,y) f(y) 
    = (\cA f)(x).
\end{align*}
\end{IEEEproof}

Next, recall from \cref{def:char-prod} that any $\chi\in\widehat{\cQ}$ may be decomposed as $\chi(x) = \chi(x_1,\ldots,x_n) = \prod_{i=1}^n \chi_i(x_i)$, where $\chi_i\in\widehat{\bbZ/q_i\bbZ}$, and $\wt(\chi) = \abs*{\mathset{1\leq i\leq n}{\chi_i\neq \1_{\widehat{\bbZ/q_i\bbZ}}}}$.

\begin{lemma}
For all $\chi\in\widehat{\cQ}$ it holds that 
$\widehat{L}(\chi) = \parenv*{\sum_{\substack{1\leq i\leq n \\ \chi_i = \1}} q_i} - n = n(\qa-1) - \sum_{\substack{1\leq i\leq n \\ \chi_i \neq \1}} q_i$.
\end{lemma}
\begin{IEEEproof}
Observe 
\begin{align*}
    \widehat{L}(\chi) 
    &= \sum_{x\in\cQ} \frac{L(x)}{\abs*{\cQ}} \overline{\chi(x)} 
    = \sum_{\substack{x\in\cQ \\ \wt(x)=1}} \overline{\chi(x)} \\
    &= \sum_{\substack{x\in\cQ \\ \wt(x)=1}} \chi(-x) 
    = \sum_{i=1}^n \sum_{\substack{x_i\in(\bbZ/q_i\bbZ) \\ x_i\neq 0}} \chi_i(x_i). \numberthis
\end{align*}
Observe that by \cref{lem:sum-of-char} 
\[
\sum_{x_i\in(\bbZ/q_i\bbZ)} \chi_i(x_i) = \begin{cases}
q_i, & \chi_i = \1; \\
0, & \text{otherwise}.
\end{cases}
\]
Since $\chi_i(0)=1$, we have 
\[
\sum_{\substack{x_i\in(\bbZ/q_i\bbZ) \\ x_i\neq 0}} \chi_i(x_i) = \begin{cases}
q_i-1, & \chi_i = \1; \\
-1, & \text{otherwise}.
\end{cases}
\]
This concludes the proof.
\end{IEEEproof}

\begin{lemma}\label{lem:even-eigenfunction}
Take $B\subseteq\cQ$. If $B$ is \emph{symmetric} (namely $B=-B$; i.e., $x\in B$ if and only if $-x\in B$ for all $x\in\cQ$) then there exists an even eigenfunction $\fB\colon\cQ\to\bbR$, $\fB\geq 0$, of $\lB$ (namely, $f(x)=f(-x)$ for all $x\in\cQ$).
\end{lemma}
\begin{IEEEproof}
Take a non-negative eigenfunction $\fB$ guaranteed by Perron's theorem. Since the Hamming metric is invariant to inversions, i.e., $d(x,y) = d(-x,-y)$, it follows that $\cA(x,y)=\cA(-x,-y)$ for all $x,y\in\cQ$.

Define a function $\fB^-\colon\cQ\to\bbR$ by $\fB^-(x)\deq \fB(-x)$. Observe that 
\begin{align*}
	(\cA\fB^-)(x) &= \sum_{y\in\cQ} \cA(x,y) \fB^-(y) 
	= \sum_{y\in\cQ} \cA(x,y) \fB(-y) \\
	&= \sum_{z\in\cQ} \cA(x,-z) \fB(z) 
	= \sum_{z\in\cQ} \cA(-x,z) \fB(z) \\
	&= (\cA\fB)(-x) = \lB \fB(-x) = \lB \fB^-(x).
\end{align*}
That is, $\fB^-$ is also a non-negative eigenfunction of $\lB$. Note, then, that $\fB^{\mathtt{e}}\deq \frac{\fB+\fB^-}{2}\geq 0$ is even, and also an eigenfunction, as required.
\end{IEEEproof}

In what follows, we utilize the \emph{characteristic function} of a code $C\subseteq\cQ$ by 
\begin{align}
    \1_C(x)\deq \begin{cases}
    1, & x\in C; \\
    0, & x\not\in C.
    \end{cases}
\end{align}

\begin{lemma}\label{lem:expectation-ratio}
Let $C\subseteq\cQ$ be a code, and define $\phi\colon\cQ\to\bbC$ by 
\[
\phi(x)\deq \sum_{y\in\cQ} \sqrt{\1_C*\1_{-C}(y)} \widehat{y}(x),
\]
where $\1_{-C}(x) = \1_{\mathset*{-y}{y\in C}}(x) = \1_C(-x)$ (i.e., 
$\1_{-C} = \1_C^-$). 
Then $\widehat{\phi}(\widehat{x}) = \sqrt{\1_C*\1_{-C}(x)}$, $\phi$ is 
even and real-valued, $\phi*\phi\geq 0$, and 
\[
\frac{\bbE(\phi^2)}{\parenv[\big]{\bbE(\phi)}^2} = \abs*{C}.
\]
\end{lemma}
\begin{IEEEproof}
First, note for all $x\in\cQ$ that $\1_C * \1_{-C} (x) = \bbE(\1_C 
\1_{-C}^x)\geq 0$, hence $\phi$ is well-defined. 
We also note by \cref{lem:even-real} that $\widehat{\1_{-C}}(\chi) = 
\widehat{\1_C}(\chi^{-1}) = \overline{\widehat{\1_C}(\chi)}$. 
Further, observe for $x\in\cQ$ that 
\begin{align*}
	\widehat{\phi}(\widehat{x}) 
	&= \frac{1}{\abs*{\cQ}} \sum_{y\in\cQ} \phi(y) \overline{\widehat{x}(y)} \\
	&= \frac{1}{\abs*{\cQ}} \sum_{y\in\cQ} 
	\parenv*{\sum_{z\in\cQ} \sqrt{\1_C*\1_{-C}(z)} \widehat{z}(y)} 
	\overline{\widehat{x}(y)} \\
	&= \sum_{z\in\cQ} \sqrt{\1_C*\1_{-C}(z)} 
	\parenv*{\frac{1}{\abs*{\cQ}} \sum_{y\in\cQ} \widehat{z}(y) \overline{\widehat{x}(y)}} \\
	&= \sum_{z\in\cQ} \sqrt{\1_C*\1_{-C}(z)} 
	\parenv*{\frac{1}{\abs*{\cQ}} \sum_{y\in\cQ} \widehat{(z-x)}(y)} \\
	&= \sqrt{\1_C*\1_{-C}(x)}.
\end{align*}
In particular, since 
\begin{align*}
	\1_C*\1_{-C}(x) 
	&= \frac{1}{\abs*{\cQ}} \sum_{y\in\cQ} \1_C(y) \1_{-C}(x-y) \\
	&= \frac{1}{\abs*{\cQ}} \sum_{y\in\cQ} \1_{-C}(-y) \1_C(-x-(-y)) \\
	&= \frac{1}{\abs*{\cQ}} \sum_{z\in\cQ} \1_{-C}(z) \1_C(-x-z) \\
	&= \1_{-C}*\1_C(-x) = \1_C*\1_{-C}(-x)
\end{align*}
is even, so is $\widehat{\phi}$ (and trivially real-valued). Hence by 
\cref{lem:even-real} we have that $\phi$ is also even and real-valued, 
and that $\widehat{(\1_C*\1_{-C})}$ is even (and real-valued).

Moreover, since $\1_C*\1_{-C}(x) = 
\parenv*{\widehat{\phi}(\widehat{x})}^2 = 
\widehat{(\phi*\phi)}(\widehat{x})$, i.e., $\widehat{(\phi*\phi)} = 
\widecheck{(\1_C*\1_{-C})}$. By \cref{lem:dual-transform}, for all 
$x\in\cQ$ we have 
\begin{align*}
	\phi*\phi(x) 
	&= \abs*{\cQ} \widehat{(\1_C*\1_{-C})}(\widehat{x}^{-1}) 
	= \abs*{\cQ} \widehat{(\1_C*\1_{-C})}(\widehat{x}) \\
	&= \abs*{\cQ} \widehat{\1_C}(\widehat{x}) 
	\widehat{\1_{-C}}(\widehat{x}) 
	= \abs*{\cQ} \abs*{\widehat{\1_C}(\widehat{x})}^2 
	\geq 0,
\end{align*}
where we again use \cref{lem:even-real}.

Next, we observe that 
\begin{align*}
    \parenv*{\bbE(\phi)}^2 
    &= \parenv[\big]{\widehat{\phi}(\1)}^2 
    = \1_C*\1_{-C}(0) \\
    &= \frac{1}{\abs*{\cQ}} \sum_{x\in\cQ} \1_C(x) \1_{-C}(-x) \\
    &= \frac{1}{\abs*{\cQ}} \sum_{x\in\cQ} \1_C(x)^2 
    = \frac{\abs*{C}}{\abs*{\cQ}}.
\end{align*}
Further, 
\begin{align*}
	\bbE\parenv*{\phi^2} 
    &= \frac{1}{\abs*{\cQ}} \sum_{x\in\cQ} \parenv*{\phi(x)}^2 
    = \frac{1}{\abs*{\cQ}} \sum_{x\in\cQ} \phi(x) \phi(-x) \\
    &= \phi*\phi(0) 
    = \abs*{\cQ} \abs*{\widehat{\1_C}(\widehat{0})}^2 
    = \frac{\abs*{C}^2}{\abs*{\cQ}}.
\end{align*}
This concludes the proof.
\end{IEEEproof}

Now, we are ready to prove \cref{th:bound-by-ev}, which we restate.
{
\renewcommand{\thetheorem}{\ref{th:bound-by-ev}}
\begin{theorem}
Let $C\subseteq\cQ$ be a code with minimum distance $d$. Take a 
symmetric $B\subseteq\cQ$ such that $\lB\geq (n+1) (\qa-1) - 
\sum_{i=1}^d q_i$. Then 
\begin{align*}
    \abs*{C}\leq n\abs*{B}.
\end{align*}
\end{theorem}
\addtocounter{theorem}{-1}
} 
\begin{IEEEproof}
We again define $\phi\colon\cQ\to\bbR$ as in 
\cref{lem:expectation-ratio}. 
Observe that $C$ has minimum distance~$d$ if and only if 
$\1_C*\1_{-C}(x) = 0$ for all $x\in\cQ$ satisfying $0<\wt(x)<d$. Hence 
$\widehat{\phi}(\chi)=0$ for all $\chi\in\widehat{\cQ}$ satisfying 
$0<\wt(\chi)<d$.

Next, recall by \cref{lem:even-eigenfunction} that there exists an even and non-negative $\fB\geq 0$, $\supp(\fB)\subseteq B$, such that $\cA\fB = \lB\fB$. Since $\cA$ and $\fB$ are non-negative, it trivially holds for $x\in\cQ\setminus B$ that $\cA\fB(x) \geq 0 = \lB\fB(x)$; for all $x\in\cQ$, then, $\cA\fB(x) \geq \lB\fB(x)$.

We define $h\colon\cQ\to\bbR$ by $h\deq \phi*\fB$. Then, observe 
\begin{align*}
	\inner*{\cA h}{h} &= \inner*{h * L}{h} 
	= \inner*{(\phi*\fB)*L}{\phi*\fB} \\
	&= \inner*{\phi*(\fB*L)}{\phi*\fB} \\
	&= \inner*{\fB*L}{\phi*(\phi*\fB)} 
	= \inner*{\cA\fB}{\phi*(\phi*\fB)} \\
	&\geq \lB\inner*{\fB}{\phi*(\phi*\fB)} 
	= \lB\inner*{\phi*\fB}{\phi*\fB} \\
	&= \lB \bbE(h^2),
\end{align*}
where the inequality follows from the fact that $\phi*(\phi*\fB) = 
(\phi*\phi)*\fB$ is non-negative (since $\phi*\phi, \fB\geq 0$) and 
$\cA\fB(x) \geq \lB\fB(x)$, and the last equality from the fact that 
$h$ is real-valued.

On the other hand, using \cref{thm:Plancharel} 
\begin{align*}
	\inner*{\cA h}{h} &= \inner*{h * L}{h} 
	= \inner*{\widehat{h} \widehat{L}}{\widehat{h}} \\
	&= \sum_{\chi\in\widehat{\cQ}} \widehat{L}(\chi) \abs*{\widehat{h}(\chi)}^2 \\
	&= \sum_{\chi\in\widehat{\cQ}} 
	\parenv[\bigg]{n(\qa-1) 
	- \sum_{\substack{1\leq i\leq n \\ \chi_i \neq \1}} q_i} 
	\abs*{\widehat{h}(\chi)}^2.
\end{align*}
Note that $\abs[\big]{\widehat{h}(\chi)}^2 = \abs[\big]{\widehat{\phi}(\chi)}^2 \abs[\big]{\widehat{\fB}(\chi)}^2 = 0$ for all $0<\wt(\chi)<d$; therefore 
\begin{IEEEeqnarray*}{+rCl+x*}
	\inner*{\cA h}{h} 
	&=& n(\qa-1) \abs*{\widehat{h}(\1)}^2 \\
	&& +\> \sum_{\substack{\chi\in\widehat{\cQ} \\ \wt(\chi)\geq d}} 
	\parenv[\bigg]{n(\qa-1) 
	- \sum_{\substack{1\leq i\leq n \\ \chi_i \neq \1}} q_i} 
	\abs*{\widehat{h}(\chi)}^2 \\
	&\leq& n(\qa-1) \abs*{\widehat{h}(\1)}^2 \\
	&& +\> \parenv[\bigg]{n(\qa-1) - \sum_{i=1}^d q_i} 
	\sum_{\chi\in\widehat{\cQ}} \abs*{\widehat{h}(\chi)}^2 \\
	&=& n(\qa-1) \parenv*{\bbE(h)}^2 
	+ \parenv[\bigg]{n(\qa-1) - \sum_{i=1}^d q_i} \bbE(h^2),
\end{IEEEeqnarray*}
where $\sum_{\chi\in\widehat{\cQ}} \abs[\big]{\widehat{h}(\chi)}^2 = \bbE(h^2)$ uses \cref{thm:Plancharel} and the fact that $h$ is real-valued.

Then, we have observed 
\begin{IEEEeqnarray*}{+rCl+x*}
	\lB \bbE(h^2) &\leq& n(\qa-1) \parenv*{\bbE(h)}^2 \\
	&& +\> \parenv[\bigg]{n(\qa-1) - \sum_{i=1}^d q_i} \bbE(h^2), 
\end{IEEEeqnarray*}
and from the assumption $\lB\geq (n+1) (\qa-1) - \sum_{i=1}^d q_i$, 
\begin{align}
	\bbE(h^2) \leq n \parenv*{\bbE(h)}^2.
\end{align}

We can now observe 
\begin{align*}
	\parenv*{\bbE(h)}^2 &= \parenv*{\bbE(\phi*\fB)}^2 
	= \parenv*{\widehat{(\phi*\fB)}(\1)}^2 \\
	&= \parenv*{\widehat{\phi}(\1) \widehat{\fB}(\1)}^2 
	= \parenv[\bigg]{\bbE(\phi) \bbE(\fB)}^2, 
\end{align*}
and 
\begin{align*}
	\bbE\parenv*{h^2} &= \inner*{h}{h} 
	= \inner*{\phi*\fB}{\phi*\fB} \\
	&= \inner*{\phi*(\phi*\fB)}{\fB} 
	= \inner*{(\phi*\phi)*\fB}{\fB} \\
	&= \inner*{\fB*(\phi*\phi)}{\fB} 
	= \inner*{\phi*\phi}{\fB*\fB} \\
	&= \frac{1}{\abs*{\cQ}} \sum_{x\in\cQ} \phi*\phi(x) \overline{\fB*\fB(x)} \\
	&= \frac{1}{\abs*{\cQ}} \sum_{x\in\cQ} \phi*\phi(x) \fB*\fB(x) \\
	&\geq \frac{1}{\abs*{\cQ}} \phi*\phi(0) \fB*\fB(0) 
	= \frac{1}{\abs*{\cQ}} \bbE(\phi^2) \bbE(\fB^2),
\end{align*}
where the inequality follows from $\phi*\phi\geq 0$ (see \cref{lem:expectation-ratio}) and $\fB*\fB\geq 0$ (since $\fB\geq 0$), and the last equality from the fact that $\phi,\fB$ are even.

Summarizing, we have seen that 
\begin{align}
	\abs*{\cQ} \frac{\parenv[\big]{\bbE(\fB)}^2}{\bbE(\fB^2)} 
	\geq \frac{1}{n} \frac{\bbE(\phi^2)}{\parenv[\big]{\bbE(\phi)}^2} 
	= \frac{1}{n} \abs*{C}.
\end{align}

Finally, since $\supp(\fB)\subseteq B$ and $\fB\geq 0$, observe from the Cauchy-Schwarz inequality that 
\begin{align*}
	\parenv[\big]{\bbE(\fB)}^2 
	&= \parenv[\big]{\inner*{\fB}{\1_B}}^2 
	\leq \bbE(\fB^2) \bbE(\1_B^2) 
	= \frac{\abs*{B}}{\abs*{\cQ}} \bbE(\fB^2),
\end{align*}
which concludes the proof.
\end{IEEEproof}

\subsection{Eigenvalues of Hamming spheres}\label{sec:ball-ev}

\begin{theorem}\label{thm:ball-eigenvalue}
Take $\sqrt{n}<r\leq n$. Then 
\begin{IEEEeqnarray*}{+rCl+x*}
	\lambda_{B_r(0)} &\geq& \frac{1}{\floorenv*{\sqrt{n}}} 
	\sum_{k=r-\floorenv*{\sqrt{n}}+2}^{r-1} 
	\Bigg(k \sqrt{\frac{s_k}{s_{k-1}}} 
	+ n (\qa-1) \phantom{\Bigg).} \\
	\IEEEeqnarraymulticol{3}{r}{-\> k - (k+1) \frac{s_{k+1}}{s_k} 
	+ (k+1) \sqrt{\frac{s_{k+1}}{s_k}}\Bigg).}
\end{IEEEeqnarray*}
\end{theorem}
\begin{IEEEproof}
For ease of notation, denote $M\deq \floorenv*{\sqrt{n}}$. We shall find $f\colon\cQ\to\bbR$, $f\geq 0$, $\supp(f)\subseteq B_r(0)$ such that $\frac{\inner*{\cA f}{f}}{\inner*{f}{f}}$ satisfies the proposition.

Indeed, define 
\begin{align*}
	f(x)\deq \begin{cases}
	\frac{1}{\sqrt{s_{\wt(x)}}}, & r-M < \wt(x)\leq r; \\
	0, & \text{otherwise}.
	\end{cases}
\end{align*}
Then  
\begin{align*}
	\inner*{f}{f} &= \sum_{k=r-M+1}^r  \sum_{\substack{x\in\cQ \\ \wt(x) = k}} \frac{1}{s_k} \\
	&= \sum_{k=r-M+1}^r 1 = M,
\end{align*}
and 
\begin{align*}
	\inner*{\cA f}{f} &= \sum_{x\in\cQ} (\cA f)(x) \overline{f(x)} \\
	&= \sum_{x\in\cQ} \sum_{\substack{y\in\cQ \\ \wt(y)=1}} f(x+y) f(x) \\
	&= \sum_{k=r-M+1}^r \sum_{\substack{x\in\cQ \\ \wt(x)=k}} f(x) 
	\sum_{i=1}^n \sum_{\substack{y\in\cQ \\ \supp(y)=\bracenv*{i}}} 
	f(x+y) \\
	&= \sum_{k=r-M+1}^r \frac{1}{\sqrt{s_k}} \sum_{i=1}^n 
	\sum_{\substack{y\in\cQ \\ \supp(y)=\bracenv*{i}}} 
	\sum_{\substack{x\in\cQ \\ \wt(x)=k}} f(x+y) \\
	&\geq \sum_{k=r-M+2}^{r-1} \frac{1}{\sqrt{s_k}} \sum_{i=1}^n 
	\sum_{\substack{y\in\cQ \\ \supp(y)=\bracenv*{i}}} 
	\sum_{\substack{x\in\cQ \\ \wt(x)=k}} f(x+y)
\end{align*}
(where the last inequality is taken so that $f(x+y)\neq 0$ for all $x,y$.)

Fixing $r-M+1 < k < r$, $1\leq i\leq n$ and $y\in\cQ$ such that $\supp(y)=\bracenv*{i}$, we find $\sum_{\substack{x\in\cQ \\ \wt(x)=k}} f(x+y)$: 
\begin{itemize}
\item For $x\in\cQ$ such that $i\in\supp(x)$ and $x_i=-y_i$ (of which there are $s_{k-1}([n]\setminus\bracenv*{i})$) the contribution is $1/\sqrt{s_{k-1}}$.

\item For $x\in\cQ$ such that $i\in\supp(x)$ and $x_i\neq -y_i$ (of which there are $(q_i-2) s_{k-1}([n]\setminus\bracenv*{i})$) the contribution is $1/\sqrt{s_k}$.

\item For $x\in\cQ$ such that $i\not\in\supp(x)=\emptyset$ (of which there are $s_k([n]\setminus\bracenv*{i})$) the contribution is $1/\sqrt{s_{k+1}}$.
\end{itemize}

We then have, for $k$ and $i$ as above 
\begin{IEEEeqnarray*}{+l+x*}
	\sum_{\substack{y\in\cQ \\ \supp(y)=\bracenv*{i}}} \sum_{\substack{x\in\cQ \\ \wt(x)=k}} f(x+y) 
	= (q_i-1) \Bigg(\frac{s_{k-1}([n]\setminus\bracenv*{i})}{\sqrt{s_{k-1}}}\phantom{\Bigg).} \\
	\IEEEeqnarraymulticol{1}{r}{+\> \frac{(q_i-2) s_{k-1}([n]\setminus\bracenv*{i})}{\sqrt{s_k}} + \frac{s_k([n]\setminus\bracenv*{i})}{\sqrt{s_{k+1}}}\Bigg).}
\end{IEEEeqnarray*}

Now, from \cref{lem:sum-sphere-not-supported}:
\begin{align*}
	\sum_{i=1}^n (q_i-1) s_{k-1}([n]\setminus\bracenv*{i}) 
	&= k s_k;
\end{align*}
similarly, 
\begin{align*}
	\sum_{i=1}^n (q_i-1) s_k([n]\setminus\bracenv*{i}) 
	&= (k+1) s_{k+1};
\end{align*}
and finally 
\begin{IEEEeqnarray*}{+rCl+x*}
	\IEEEeqnarraymulticol{3}{l}{\sum_{i=1}^n (q_i-1) (q_i-2) s_{k-1}([n]\setminus\bracenv*{i})} \\
	\;\; &=& \sum_{i=1}^n (q_i-1)^2 s_{k-1}([n]\setminus\bracenv*{i}) \\
	&& -\> \sum_{i=1}^n (q_i-1) s_{k-1}([n]\setminus\bracenv*{i}) \\
	&=& n(\qa-1) s_k - (k+1) s_{k+1} - k s_k.
\end{IEEEeqnarray*}
Summarizing, 
\begin{IEEEeqnarray*}{+rCl+x*}
	\IEEEeqnarraymulticol{3}{l}{\frac{1}{\sqrt{s_k}} \sum_{i=1}^n 
	\sum_{\substack{y\in\cQ \\ \supp(y)=\bracenv*{i}}} 
	\sum_{\substack{x\in\cQ \\ \wt(x)=k}} f(x+y)} \\
	\;\; &=& k \sqrt{\frac{s_k}{s_{k-1}}} 
	+ n (\qa-1) - k - (k+1) \frac{s_{k+1}}{s_k} \phantom{,} \\
	\IEEEeqnarraymulticol{3}{r}{+\> (k+1) \sqrt{\frac{s_{k+1}}{s_k}},}
\end{IEEEeqnarray*}
which concludes the proof.
\end{IEEEproof}

Finally, \cref{cor:ball-eigenvalue} follows from the last theorem. 
Again, we shall restate it before the proof.
{
\renewcommand{\thecorollary}{\ref{cor:ball-eigenvalue}}
\begin{corollary}
Take $2\sqrt{n}<r\leq n$. Then $\lambda_{B_r(0)} \geq 
2\sqrt{(\qa-1) r (n-r)} + (\qa-2) r + o(n)$.
\end{corollary}
\addtocounter{corollary}{-1}
} 
\begin{IEEEproof}
We again denote $M\deq \floorenv*{\sqrt{n}}$. 
First, note by \cref{thm:sphere-ratio-ave} 
that for all $M\leq k<n$ (and, indeed, for all $r-M+2\leq k<r$) 
\begin{align*}
	k \sqrt{\frac{s_k}{s_{k-1}}} 
	&\geq (k+1) \sqrt{\frac{s_{k+1}}{s_k}}\cdot 
	\sqrt{\frac{(n-k+1) k}{(n-k) (k+1)}} \\
	&= (k+1) \sqrt{\frac{s_{k+1}}{s_k}}\cdot 
	\sqrt{\frac{1 + 1/(n-k)}{1 + 1/k}} \\
	&\geq (k+1) \sqrt{\frac{s_{k+1}}{s_k}}\cdot 
	\sqrt{\frac{1 + 1/(n-M)}{1 + 1/M}}.
\end{align*}
Therefore, by \cref{thm:ball-eigenvalue} 
\begin{IEEEeqnarray*}{+l+x*}
	\lambda_{B_r(0)} \geq \frac{1}{M} 
	\sum_{k=r-M+2}^{r-1} 
	\Bigg(n (\qa-1) - k \\
	\IEEEeqnarraymulticol{1}{r}{-\> (k+1) \sqrt{\frac{s_{k+1}}{s_k}} 
	\parenv*{\sqrt{\frac{s_{k+1}}{s_k}} - 1 
	- \sqrt{\frac{1 + 1/(n-M)}{1 + 1/M}}}\Bigg).}
\end{IEEEeqnarray*}
We denote $\alpha_n\deq 1 + \sqrt{\frac{1 + 1/(n-M)}{1 + 1/M}}$, and 
note that $\alpha_n = 2+o(1)$. Applying \cref{thm:sphere-ratio-ave} 
we now obtain 
\begin{IEEEeqnarray*}{+l+x*}
	\lambda_{B_r(0)} \geq \frac{1}{M} 
	\sum_{k=r-M+2}^{r-1} 
	\Big((\qa-2) k \phantom{\Big)} \\
	\IEEEeqnarraymulticol{1}{r}{+\> \alpha_n \sqrt{(\qa-1) (k+1) (n-k)}\Big)}
\end{IEEEeqnarray*}

Next, observe for all $r-M+1<k<r$ that 
\begin{align*}
	\sqrt{(k+1) (n-k)} 
	&> \sqrt{(r-M) (n-r)} \\
	&> \sqrt{r (n-r)} - \sqrt{M (n-r)} \\
	&> \sqrt{r (n-r)} - n^{3/4}.
\end{align*}
Also, 
\begin{align*}
	\sum_{k=r-M+2}^{r-1} k 
	&= (M-2) \parenv*{r - \frac{M-1}{2}}.
\end{align*}

Finally, we've seen that 
\begin{IEEEeqnarray*}{+rCl+x*}
	\lambda_{B_r(0)} &\geq& 
	\frac{M-2}{M} \Bigg(\alpha_n \sqrt{\qa-1} 
	\parenv*{\sqrt{r (n-r)} - n^{3/4}} \phantom{\Bigg)} \\
	\IEEEeqnarraymulticol{3}{r}{+\> (\qa-2) \parenv*{r - \frac{M-1}{2}}\Bigg)} \\
	&=& (2+o(1))\sqrt{(\qa-1) r (n-r)} + (\qa-2) r \phantom{).} \\
	\IEEEeqnarraymulticol{3}{r}{-\> O(n^{3/4}).}
	\\[-\normalbaselineskip] &&&\IEEEQEDhere
\end{IEEEeqnarray*}
\end{IEEEproof}


%









\end{document}